\documentclass[12pt,english,preprint]{revtex4-1}

\usepackage{times}
\usepackage{array}
\usepackage{float}
\usepackage{placeins}
\usepackage{amsmath}
\usepackage{mathrsfs}
\usepackage{graphicx}
\usepackage{amssymb}
\usepackage{wrapfig}
\usepackage{color}
\usepackage[FIGTOPCAP]{subfigure}
\usepackage{bm}
\newcommand{\fracxi}[1]{\frac{\partial #1}{\partial \xi}}
\newcommand{\fracp}[1]{\frac{\partial #1}{\partial p}}

\newcommand{\fract}[1]{\frac{\partial #1}{\partial t}}
\makeatletter
\usepackage{fancyhdr}
\usepackage{babel}
\makeatother
\usepackage[colorlinks=true,linkcolor=blue,citecolor=blue,urlcolor=blue]{hyperref}
\hypersetup{draft=false}

\begin{document}


\title{Hierarchical Framework of Runaway Electrons using Deep Learning}

\author{Tyler Mark}
\email{tyler.mark@ufl.edu}
\author{Christopher J. McDevitt}
\email{cmcdevitt@ufl.edu}
\affiliation{Nuclear Engineering Program, Department of Materials Science and Engineering, University of Florida, Gainesville, FL 32611, United States of America}


\date{\today}

\begin{abstract}


We present an adjoint deep learning framework describing the evolution of fluid moments and the energy distribution of the runaway electron (RE) population. 
We demonstrate that a careful formulation of the adjoint problem allows for the temporal evolution of these quantities for arbitrary initial electron distributions,
and in combination with a physics-informed neural network (PINN), we show that the resulting surrogates can resolve a broad range of plasma parameters. 
This combination of the adjoint formulation and rapid inference of neural networks enables orders of magnitude faster predictions of RE kinetics than traditional methods.
Here, we detail the mathematical formulation and the design of three PINNs which recover the temporal evolution of the RE current, average energy and energy distribution.
Predictions are validated against a traditional RE solver, with good agreement across a broad range of scenarios.

\end{abstract}

\maketitle

\section{Introduction}
\label{sec:introduction}
The description of runaway electron (RE) formation and evolution in magnetized fusion devices poses a substantial scientific challenge, particularly when coupling their dynamics with magnetohydrodynamic (MHD) models \cite{bandaru_hoelzl_reux_ficker_silburn_lehnen_eidietis_team_2021, liu_zhao_jardin_ferraro_paz-soldan_liu_lyons_2021, sainterme_sovinec_2024}.  
While first-principles kinetic RE models are available \cite{harvey_chan_chiu_evans_rosenbluth_whyte_2000, nilsson_decker_y_peysson_granetz_f_saint-laurent_m_vlainic_2015, mcdevitt_guo_tang_2019, hoppe_embreus_fulop_2021, beidler_del-castillo-negrete_shiraki_baylor_hollmann_lasnier_2024}, such models are computationally intensive, thus sharply limiting their application for multi-physics coupling. 
This has led to the development of several reduced fluid models \cite{zhao_liu_jardin_ferraro_2020,Jorek_runaway_extension} that may be more easily coupled to the background plasma evolution. However, these models discard crucial information regarding the energy and pitch-angle distribution of REs, reducing the physics fidelity of the resulting integrated model.

The present work aims to address these limitations by employing an adjoint deep learning framework that can directly predict the evolution of \textit{arbitrary RE moments}
and the RE energy distribution for any initial condition.
Prediction of higher-order moments typically requires evolving the full distribution either through a continuum solver for the relativistic Fokker-Planck equation or a particle-based RE solver. 
Noting the requirements of resolving a strongly anisotropic RE distribution, together with the broad range of energies characteristic of REs, 
continuum solvers are often computationally intensive. Particle-based approaches require the evolution of a large number of marker particles, resulting in substantial computational expense. 
Adjoint methods \cite{giles_pierce_2000,adjoint_equations_in_cfd,bui-thanh_2023} offer a promising approach to accelerate the evaluation of quantities of interest of the RE population. Specifically, a single adjoint solution enables the density moment of the RE population to be evolved beginning from an arbitrary initial momentum space distribution \cite{mcdevitt_arnaud_tang_2025, mcdevitt_arnaud_tang_2025_secondary}. Furthermore, once the adjoint solution is available, the RE density may be inferred at any time over which the solution to the adjoint equation is available, thus providing a rapid means of inferring how distinct initial conditions impact RE density evolution without the need for multiple forward solves of the relativistic Fokker-Planck equation.

Deep learning methods offer an avenue through which adjoint methods can be used to develop practical surrogates of RE evolution. 
Specifically, a single deep learning model can be trained across a broad range of plasma conditions, resulting in a large one-time offline training cost, but an efficient online surrogate.
The present work seeks to utilize a physics-informed neural network (PINN) \cite{raissi_perdikaris_karniadakis_2019, Karniadakis_2021} to provide a parametric solution of
the adjoint to the relativistic Fokker-Planck equation. This approach embeds physics in the training of a neural network, and while it can be used in conjunction with data, we will consider the data-free limit, with data only used to validate predictions of the model. Such a limit enables PINNs to be quickly adapted to different physical systems, containing distinct collisional coefficients for example, without the need to generate large quantities of simulation or experimental data.
We demonstrate the capability of the adjoint deep learning framework by predicting several fluid moments and the energy distribution of REs through careful treatment of the adjoint solution and benchmark against a high-fidelity Monte Carlo simulation. While specific moments necessitate tailored neural network architectures, each model generalizes across arbitrary initial conditions and a broad range of physical parameters. The resulting models provide a means of not only predicting the RE current, the quantity that directly couples to the MHD evolution, but also the average energy of the RE distribution that are critical for diagnosing and mitigating RE populations in tokamak devices. While the present study does not contain the complete set of physics necessary to describe REs under the complex conditions that emerge during startup and disruption scenarios in tokamak devices, we anticipate that the present results can be straightforwardly generalized to a more comprehensive collision operator and treat a broader range of physics parameters, similar to recent work on the RE avalanche \cite{arnaud2025runaway}. The extension of the present results to specific RE generation scenarios in tokamak devices will be the focus of future work.


The remainder of the paper is organized as follows. Section \ref{sec:PRE} describes an adjoint formulation for predicting moments of the RE distribution function together with the energy distribution. 
A brief description of the physics-constrained machine learning approach is given in Sec. \ref{sec:PINNs} along with its adaptation for the adjoint of the relativistic Fokker-Planck equation. 
Section \ref{sec:moments_results} gives predictions of the evolution of several moments of the RE distribution. 
The extension of the adjoint framework to evolve the energy distribution of REs is illustrated in Sec. \ref{sec:energy_distrib}. 
A brief discussion and conclusions are given in Sec. \ref{sec:discussion_conclusion}.
Upon acceptance of this manuscript, the version of the PINN and validation scripts used in this manuscript are available on github.com/cmcdevitt2/RunAwayPINNs.

\section{Predicting Runaway Electron Moments through the Relativistic Fokker-Planck Equation}
\label{sec:PRE}

\subsection{Relativistic Fokker-Planck Equation}
\label{sec:adjointFP}

Predicting higher order fluid moments of a distribution typically requires knowledge of the full distribution of runaways $f_{RE}$ itself. Specifically, a given moment $M (t)$ of the RE distribution can be written:
\begin{equation}
    M(t) = \int d^3p g(p,\xi) f_{RE}(p,\xi,t)
    , \label{eq:Moment_full_distribution}
\end{equation}
where $d^3p=2\pi p^2 dp d\xi$ and $g(p,\xi)$ is a weighting term defining the moment of the distribution [$g (p,\xi) = 1$ for density, for example]. This section will demonstrate that the adjoint formulation can evaluate any moment of the 
runaway distribution without knowledge of the full distribution $f_{RE}$. We begin by noting that the Green's function $F(\mathbf{p},t;\mathbf{p_0},t_0)$
allows the runaway distribution to be propagated from a given initial condition $f_{init}(\mathbf{p_0},t_{0})$ via the integral \cite{hannes_risken_2019},
\begin{gather}
    f_{RE}(\mathbf{p},t)=\int d^3p_0 f_{init}(\mathbf{p_0},t_0) F(\mathbf{p},t;\mathbf{p_0},t_0),
    \label{eq:Full Distribution Green's function}
\end{gather}
where $\mathbf{p}$ indicates momentum space coordinates $(p,\xi)$ and the integration is over the initial condition coordinates $\mathbf{p_0}$.
Substituting Eq. (\ref{eq:Full Distribution Green's function}) into (\ref{eq:Moment_full_distribution}) yields an expression for any RE moment $M (t)$ in terms of the Green's function $F (\mathbf{p},t;\mathbf{p_0},t_0)$, i.e.,
\begin{equation}
    M(t) = \int \, d^3p_0 f_{init}(\mathbf{p_0},t_0) \int  d^3p g(p,\xi) F(\mathbf{p},t;\mathbf{p_0},t_0).
    \label{eq:moment_1}
\end{equation}
Thus, by computing the Green's function, an arbitrary moment of the RE distribution can be inferred after carrying out integrals over $\mathbf{p}$ and $\mathbf{p_0}$. However, computing the Green's function is challenging since it depends on both $\mathbf{p}$ and $\mathbf{p}_0$, where the computational burden of evaluating the double integration appearing in Eq. (\ref{eq:moment_1}) provides a further obstacle to this approach. In the following, we demonstrate that an adjoint formulation can provide a more rapid means of estimating moments of the RE distribution within a finite domain for any initial condition.
We begin by noting that the Green's function of the relativistic Fokker-Planck equation obeys
\begin{subequations}
\begin{gather}
    \frac{\partial F(\mathbf{p},t;\mathbf{p_0},t_0)}{\partial t}+L(F)= \delta(\mathbf{p-p_0})\delta(t-t_0)
    , \label{eq:KineticOperators}
\end{gather}
where $L(F)$ is a sum of operators,
\begin{eqnarray}
    L(F) \equiv E(F)+ C(F)+R(F),
\end{eqnarray} 
\end{subequations}
such that the Green's function evolves due to electric field acceleration $E(F)$, small-angle collisions $C(F)$, and losses due to synchrotron radiation $R(F)$, whose forms are defined by \cite{guo_mcdevitt_tang_2017}:
\begin{subequations}
\begin{equation}
    E(F)= -\frac{1}{p^2}\frac{\partial}{\partial p}\left[p^2E_\Vert \xi F(\mathbf{p},t;\mathbf{p_0},t_0)\right]-\frac{\partial}{\partial \xi}\left[\left(\frac{1-\xi^2}{p}\right)E_\Vert F(\mathbf{p},t;\mathbf{p_0},t_0)\right]
    , \label{eq:E}
\end{equation}
\begin{equation}
    C(F)= -\frac{1}{p^2}\frac{\partial}{\partial p}p^2\left[C_F F(\mathbf{p},t;\mathbf{p_0},t_0)\right] - \frac{C_B}{p^2}\frac{\partial}{\partial \xi} \left[(1-\xi^2)\frac{\partial F(\mathbf{p},t;\mathbf{p_0},t_0)}{\partial \xi}\right]
    ,
\end{equation}
\begin{equation}
    R(F)= -\frac{1}{p^2}\frac{\partial}{\partial p}\left[\alpha p^3 \gamma (1-\xi^2)F(\mathbf{p},t;\mathbf{p_0},t_0)\right]+\frac{\partial}{\partial \xi}\left[\alpha \frac{\xi(1-\xi^2)}{\gamma}F(\mathbf{p},t;\mathbf{p_0},t_0)\right]
    .
\end{equation}
\end{subequations}
Here, the coefficients for collisional drag $C_F$ and pitch-angle scattering $C_B$ are defined by:
\begin{gather}
    C_F \equiv \frac{\gamma^2}{p^2},  \indent C_B \equiv \frac{\gamma}{2}\frac{\left(Z_{eff}+1\right)}{p},
\end{gather}
where the relativistic momentum is normalized to $p\rightarrow p/(m_e c)$, the Lorentz factor is $\gamma = \sqrt{1+p^2}$, the electron's pitch is defined as $\xi\equiv p_\parallel/p$, 
time is normalized to the relativistic collision time $t\rightarrow t/\tau_c$ where $\tau_c\equiv 4\pi \epsilon_0^2 m_e^2 c^3/(e^4 n_e \ln\Lambda)$,
the parallel electric field is normalized to the Connor-Hastie electric field \cite{connor_hastie_1975}
$E_\parallel \rightarrow E_\parallel/E_c$ where $E_c\equiv m_ec/(e\tau_c)$,  and the radiative strength $\alpha\equiv\tau_c/\tau_s$ is defined as a ratio of the collision timescale to the synchrotron timescale $\tau_s= 6\pi \epsilon_0 m^3_e c^3/ (e^4B^2)$.

\subsection{Defining the adjoint solution}
\label{sec:adjoint_solution}
An efficient approach to infer quantities of interest involves formulating an adjoint problem, such as was done for wave-driven currents \cite{karney_fisch_1986, antonsen_chu_1982, taguchi_1983} 
and applied to the study of the runaway probability function (RPF) \cite{karney_fisch_1986, liu_brennan_boozer_bhattacharjee_2016, arnaud_mark_mcdevitt_2024}. The adjoint of the relativistic Fokker-Planck equation can be expressed as:
\begin{subequations}
\begin{gather}
    \frac{\partial G}{\partial t}-L^*(G)=0,
    \label{eq:adjointFP1}
\end{gather}
\begin{eqnarray}
    L^*(G) = E^*(G) + C^*(G) + R^*(G),
\end{eqnarray}
\end{subequations}
where the adjoint operators are found through successive integration by parts
\begin{equation}
    \int d^3p \left[GL(F)- FL^*(G)\right] = 
    2\pi \int d\xi [p^2 U_p F G]_{p_{\min}}^{p_{\max}},
    \label{eq:adjoint_relation}
\end{equation}
where the energy flow of REs, $U_p$, is a balance between electric field acceleration, drag due to collisions and radiative losses:
\begin{eqnarray}
    U_p = -E_\Vert \xi - C_F-\alpha p \gamma \left(1-\xi^2\right)
    . \label{Up_equation}
\end{eqnarray} 
The adjoint operators that satisfy Eq. (\ref{eq:adjoint_relation}) can be written as \cite{liu_brennan_boozer_bhattacharjee_2016, karney_fisch_1986, zhang_del-castillo-negrete_2017, mcdevitt_arnaud_tang_2025}:
\begin{subequations}
\label{AdjointRelationFP}
\begin{equation}
    E^*(G)= E_\parallel\left[\xi\fracp{G}+\frac{(1-\xi^2)}{p}\fracxi{G}\right]
    ,
\end{equation}
\begin{equation}
    C^*(G)= C_F\fracp{G}-\frac{C_B}{p^2}\frac{\partial}{\partial \xi}\left[(1-\xi^2)\fracxi{G}\right]
    ,
\end{equation}
\begin{equation}
    R^*(G)= \alpha\gamma p\left(1-\xi^2\right)\fracp{G}-\alpha\xi\frac{\left(1-\xi^2\right)}{\gamma}\fracxi{G}.
\end{equation}
\end{subequations}

Here, we define the adjoint solution by selecting terminal and boundary conditions to characterize quantities of interest for runaways. 
Substituting the operators from Eqs. (\ref{eq:KineticOperators}) and (\ref{eq:adjointFP1}) into (\ref{eq:adjoint_relation}) and integrating over time yields:
\begin{gather}
    \int_{-\infty}^{t_{final}} \, dt\int d^3p\left[G \delta(\mathbf{p-p_0})\delta(t-t_0)- G\fract{F}- F\fract{G}\right] = 2\pi \int_{-\infty}^{t_{final}} dt \int d\xi [p^2 U_p F G]_{p_{\min}}^{p_{\max}}.
\end{gather}
The Dirac delta function can be integrated directly, while the time derivative terms are combined using the product rule. 
Performing the time integration and noting that the Green's function is sourced at $t=t_0$, the lower bounds of the time integral vanish due to $F(t=-\infty)=0$, so we can write:
\begin{align}
    G(\mathbf{p_0},t_0;t_{final}) &= \int d^3p G(\mathbf{p}, t_{final})F(\mathbf{p}, t_{final} ; \mathbf{p_0}, t_0)  \nonumber \\
    & + 2\pi\int_{-\infty}^{t_{final}} dt\int d\xi \left[p^2 U_p F(\mathbf{p}, t; \mathbf{p_0}, t_0)G(\mathbf{p}, t) \right]_{p_{\min}}^{p_{\max}}.
    \label{RunawayAdjointRelationGeneral}
\end{align}
The physical meaning of the adjoint solution $G$ is determined by the choice of the terminal condition $G(\mathbf{p},t=t_{final})$ and the upper and lower energy boundary conditions. 
The best-known example is obtained by characterizing runaways via the terminal condition such that electrons above a threshold energy $p_{RE}$ are considered to be runaways and by taking $G(\mathbf{p}, t_{final}) = \Theta(p-p_{RE})$ where $\Theta(p-p_{RE})$ is a Heaviside function. 
The lower energy boundary $p_{min}$ is taken to be sufficiently low that electrons reaching $p_{min}$ will further decelerate via collisions and therefore not be runaways so we require $G(p_{min},\xi,t)=0$.
To ensure that no inflow of electrons through the upper momentum boundary is present we will require $F$ to vanish at $p_{\max}$ when $U_p<0$ and to capture runaways which accelerate through $p_{max}$ we require $G(p_{max},t)=1$ when $U_p>0$.
With these choices, Eq. (\ref{RunawayAdjointRelationGeneral}) can be written as:

\begin{align}
    P (\mathbf{p_0},t_0;t_{final}) &= \int\limits_{p>p_{RE}}d^3p F(\mathbf{p}, t_{final} ; \mathbf{p_0}, t_0) \nonumber \\
    & + 2\pi p^2_{max} \int_{-\infty}^{t_{final}} dt \int\limits_{U_p>0} d\xi U_p (p_{max}) F(p_{max}, \xi, t; \mathbf{p_0}, t_0),
    \label{RunawayAdjointRelationGeneral2}
\end{align}
where $P (\mathbf{p_0},t_0;t_{final})$ is referred to as the runaway probability function~\cite{karney_fisch_1986}. Noting that the Green's function $F(\mathbf{p}, t_{final} ; \mathbf{p_0}, t_0)$ was generated by a unit source, Eq. (\ref{RunawayAdjointRelationGeneral2}) indicates the probability that a particle remains in the energy domain, but with an energy $p>p_{RE}$ (first term on the right hand side), or that it was lost to the high energy boundary (second term on the right hand side). If $p_{RE}$ is chosen to be the value above which an electron is considered to be a RE, then the quantity $P (\mathbf{p_0},t_0;t_{final})$ can be recognized to be the probability of an electron with an initial momentum $\mathbf{p}_0$, born at time $t_0$, running away by time $t_{final}$, hence its designation as a runaway probability function~\cite{karney_fisch_1986}. As discussed in Sec. \ref{sec:TAS} below, by choosing different terminal and boundary conditions, this will allow us to define adjoint problems describing arbitrary moments of the RE distribution.

\subsection{\label{sec:TAS}The adjoint solution for runaway moments}
The quantities of interest tied to the adjoint solution can be identified by multiplying Eq. (\ref{RunawayAdjointRelationGeneral}) by the initial distribution of electrons $f_{init} (\mathbf{p_0})$ and integrating over $\mathbf{p_0}$, yielding 
\begin{align}
    \int d^3 p_0 &f_{init} (\mathbf{p_0}) G(\mathbf{p_0},t_0;t_{final}) \nonumber \\
    &= \int d^3 p_0 f_{init} (\mathbf{p_0}) \int d^3p G(\mathbf{p}, t_{final})F(\mathbf{p}, t_{final} ; \mathbf{p_0}, t_0) \nonumber \\
    & + 2\pi p^2_{max} \int d^3 p_0 f_{init} (\mathbf{p_0}) \int_{-\infty}^{t_{final}} dt \int\limits_{U_p>0} d \xi U_p (p_{max}, \xi) F(p_{max}, \xi, t; \mathbf{p_0}, t_0)G(p_{max}, \xi , t)
    . \label{eq:TAS1}
\end{align}
Interchanging the order of integration for momentum-space integrals on the right-hand side (RHS) yields
\begin{align}
    \int d^3 p_0 &f_{init} (\mathbf{p_0}) G(\mathbf{p_0},t_0;t_{final}) \nonumber \\
    &= \int d^3p G(\mathbf{p}, t_{final}) \int d^3p_0  f_{init} (\mathbf{p_0})F(\mathbf{p}, t_{final} ; \mathbf{p_0}, t_0) \nonumber \\
    & + 2\pi p^2_{max} \int_{-\infty}^{t_{final}} dt \int\limits_{U_p>0} d \xi  U_p (p_{max}, \xi) G(p_{max}, \xi , t) \int d^3 p_0 f_{init} (\mathbf{p_0}) F(p_{max}, \xi, t; \mathbf{p_0}, t_0)
    . \label{eq:TAS2}
\end{align}
Now noting Eq. (\ref{eq:Full Distribution Green's function}), we can express the integrals over $d^3p_0$ in terms of the electron distribution, giving
\begin{align}
    \int d^3 p_0 f_{init} (\mathbf{p_0}) G(\mathbf{p_0},t_0;t_{final})&= \int d^3 p G(\mathbf{p}, t_{final}) f_e (\mathbf{p}, t_{final}) \nonumber \\
    & + 2\pi p^2_{max} \int_{-\infty}^{t_{final}} dt \int\limits_{U_p>0} d \xi U_p (p_{max}, \xi ,t) G(p_{max}, \xi , t) f_e(p_{max},\xi,t)
    , \label{eq:TAS3}
\end{align}
Equation (\ref{eq:TAS3}) provides the key result for defining the physical meaning of distinct adjoint problems. Specifically, the first term on the RHS is simply the moment of the terminal condition $G(\mathbf{p}, t_{final})$ over the electron distribution function at $t_{final}$. Thus, by choosing $G(\mathbf{p}, t_{final})$ to be a specific power of energy and pitch, distinct moments can be tracked. The second term is related to the time integrated momentum space flux of the quantity $G (p_{\max}, \xi , t)$ through the upper energy boundary. By choosing a consistent high energy boundary condition, this surface term will track losses of a given moment through the upper boundary of the system. An adjoint problem can thus be established by selecting a set of terminal and boundary conditions, and then integrating backward from $t_{final}$ to $t_0$, with the solution $G(\mathbf{p_0},t_0;t_{final})$ at $t_0$ allowing the value of a given RE moment to be inferred.


The simplest example can be derived by defining terminal and boundary conditions consistent with the calculation of the RPF, i.e. $G(\mathbf{p}, t_{final}) = \Theta(p-p_{RE})$ and $G(p_{max}, \xi , t) = 1$. With these choices, Eq. (\ref{eq:TAS3}) becomes
\begin{align}
    \int d^3 p_0 f_{init} (\mathbf{p_0}) P(\mathbf{p_0},t_0;t_{final})&= \int\limits_{p>p_{RE}} d^3 p f_e (\mathbf{p}, t_{final}) \nonumber \\
    & + 2\pi p^2_{max} \int_{-\infty}^{t_{final}} dt \int\limits_{U_p>0} d \xi  U_p (p_{max}, \xi ,t) f_e(p_{max},t)
    , \label{eq:TAS4}
\end{align}
such that the first term on the RHS is the number of electrons with $p>p_{RE}$ at $t_{final}$ and the second term is the time integrated flux of electrons through the high energy boundary. It is thus evident that the left hand side of Eq. (\ref{eq:TAS4}) describes the number density of REs at $t_{final}$, i.e.
\begin{equation}
	n_{RE} \left( t_{final}\right) = \int d^3p_0 f_{init} (\mathbf{p_0}) P (\mathbf{p_0},\tau),
	\label{eq:TAS5}
\end{equation}
where $\tau = t_{final}-t_0$ describes the time reversal of the RPF and the resulting number density is evaluated at the final time. 
Equation (\ref{eq:TAS5}) thus provides a description of the evolution of the number density of REs.

By taking the terminal and boundary conditions to be different functions of pitch and momentum we are able to define adjoint problems for arbitrary moments of the RE distribution. For this paper we will focus on the parallel momentum $u_\Vert$ 
and average energy $\mathscr{E}_{RE}$. Considering the parallel momentum first, a quantity directly related to the current carried by REs, we take the terminal condition $G(t=t_{final})= -v\xi\Theta(p-p_{RE})$ and the high energy boundary condition to be $G(p_{\max},U_p>0)=-v\xi$ where $v= p/\gamma$, such that Eq. (\ref{eq:TAS3}) becomes
\begin{align}
    \int d^3 p_0 f_{init} (\mathbf{p_0}) C(\mathbf{p_0},t_0;t_{final})&= -\int\limits_{p>p_{RE}} d^3 p v \xi f_e (\mathbf{p}, t_{final}) \nonumber \\
    & - 2\pi p^2_{max} \int_{-\infty}^{t_{final}} dt \int\limits_{U_p>0} d \xi U_p (p_{max}, \xi ,t) v_{max} \xi f_e(p_{max},\xi,t)
    , \label{eq:TAS6}
\end{align}
where $C (\mathbf{p_0},t_0;t_{final})$ is the current probability function (CPF) and the negative sign is due to the electron charge. The first term on the RHS of Eq. (\ref{eq:TAS6}) thus indicates the (negative) parallel momentum carried by the RE distribution between $p_{RE}$ and $p_{max}$, with the second term on the RHS indicating the parallel momentum carried through the high energy boundary. For a sufficiently large $p_{max}$, such that this high energy boundary term is negligible we have
\begin{equation}
	j_{RE} \left( t_{final}\right) = ec \int d^3p_0 f_{init} (\mathbf{p_0}) C (\mathbf{p_0},\tau)
	. \label{eq:TAS7}
\end{equation}
The adjoint problem with these modified terminal and boundary conditions thus provides a direct means of inferring the RE current, a quantity of immediate interest for closing the MHD equations. A subtlety that emerges for this moment, however, is that since electrons that exit through the high energy boundary $p_{max}$ will continue to be accelerated and have their pitch further collapsed to the $\xi=-1$ boundary, these electrons will continue to increase the magnitude of the parallel momentum they carry. While for the RE parallel current moment this will be a modest effect, since the electrons cannot move faster than the speed of light, for higher order moments this boundary term will become increasingly more relevant. This subtlety will be discussed in detail in Sec. \ref{sec:REM} below.

Considering the average relativistic kinetic energy moment of runaways, a quantity critical for mitigating wall damage,
we take the terminal condition to be $G(t=t_{final})= m_ec^2 (\gamma-1) \Theta(p-p_{RE})$ and the high energy boundary condition to be $G(p_{\max},U_p>0) = m_e c^2(\gamma-1)$, 
such that Eq. (\ref{eq:TAS3}) becomes
\begin{align}
    \int d^3 p_0 f_{init} (\mathbf{p_0}) \mathcal{E}(\mathbf{p_0},t_0;t_{final})&= m_ec^2\int\limits_{p>p_{RE}} d^3 (\gamma-1) f_e (\mathbf{p}, t_{final}) \nonumber \\
    & + 2\pi p^2_{max} m_ec^2  \int_{-\infty}^{t_{final}} dt \int\limits_{U_p>0} d \xi U_p (p_{\max}, \xi ,t) (\gamma_{\max}-1) f_e(p_{\max},\xi,t)
    , \label{eq:TAS7}
\end{align}
where $\mathcal{E}(\mathbf{p_0},t_0;t_{final})$ is the energy probability function (EPF), $m_ec^2$ is the rest mass energy of an electron equivalent to $0.511$ MeV, and $\gamma$ is the Lorentz factor where $\gamma=\sqrt{1+p^2}$. 
The first term on the RHS of Eq. (\ref{eq:TAS7}) thus indicates the kinetic energy carried by the RE distribution between $p_{RE}$ and $p_{\max}$, with 
the second term on the RHS indicates the energy of REs through the high energy boundary
whereby REs which pass through the high energy boundary are assumed to be fixed in energy at the boundary.
For a sufficiently large energy domain in which most REs are captured, the adjoint method can approximate the average energy,
\begin{align}
    \mathscr{E}(t_{final}) = \int d^3p_0 f_{init}(\mathbf{p_0})\mathcal{E}(\mathbf{p_0},\tau),
\end{align}
however, for plasma parameters in which a significant amount of REs can 
exit the domain the adjoint method will underestimate the energy due to the range of the adjoint solution maximized at the high energy boundary.
In Sec. \ref{sec:REM}, we will evolve the energy moment of the RE distribution using the adjoint-PINN method and compare with first principles solutions from a Monte Carlo RE solver to demonstrate the effects of the second term on the RHS of Eq. (\ref{eq:TAS7}).




\subsection{The Runaway Energy Distribution}
\label{sec:energy_distribution1}

A unique feature of using PINNs to solve the adjoint problems defined above is that it enables inference of the energy distribution of REs with only modest added computational cost.
Beginning by noting that the number of REs with energies greater than $p_{RE}$ is given by
\begin{eqnarray}
    n_{RE}(p_{RE},t)=\int P(p_0,\xi_0,p_{RE},\tau) f_{init}(p_0,\xi_0) d^3p_0.
    \label{adjoint_density_moment_pRE}
\end{eqnarray}
The number of electrons with energies between $p_{RE}$ and $p_{RE}+\Delta p_{RE}$, with $\Delta p_{RE}$ representing a small change to the definition of when an electron is treated as a runaway, is then given by
\begin{equation}
    \Delta n_{RE} = n_{RE}(p_{RE})-n_{RE}(p_{RE}+\Delta p_{RE})
    . \label{eq:DeltaNRE}
\end{equation}
By solving an adjoint problem with two slightly different definitions of the runaway threshold, $p_{RE}$ versus $p_{RE}+\Delta p_{RE}$, we are thus able to infer how many electrons are present in a narrow energy window. Noting that PINNs provide a convenient means of learning parametric solutions to PDEs~\cite{sun2020surrogate, mcdevitt2024physics}, we can train a single PINN to solve the adjoint problem for runaway density as a function of $p_{RE}$. Such a PINN then provides a means of not just predicting runaway density, but the number of runaways in any energy interval. This thus allows the energy distribution of runaways to be conveniently inferred, with inference cost scaling with the desired number of energy bins. As each forward prediction is rapid, this provides a highly efficient means of predicting the runaway energy distribution once the parametric PINN has been trained. 

To predict the energy distribution at a specific value we can
divide Eq. (\ref{eq:DeltaNRE}) by the volume element $2\pi p^2_{RE} \Delta p_{RE}$, yielding
\begin{equation}
    f_{RE}(p_{RE})\equiv \frac{\Delta n_{RE}}{2\pi p_{RE}^2 \Delta p_{RE}}= \frac{1}{2\pi p_{RE}^2}\frac{n_{RE}(p_{RE})-n_{RE}(p_{RE}+\Delta p_{RE})}{\Delta p_{RE}}=\frac{-1}{2\pi p_{RE}^2}\frac{\partial n_{RE}(p_{RE})}{\partial p_{RE}}
    , \label{eq:fRE}
\end{equation}
where we have taken $\Delta p_{RE}\to 0$ and noted the definition of a derivative in the final equality. Substituting Eq. (\ref{adjoint_density_moment_pRE}) into (\ref{eq:fRE}) results in an expression for the energy distribution, i.e.
\begin{eqnarray}
    f_{RE}(p_{RE},t)=-\frac{1}{2\pi p_{RE}^2}\int f_e^{init}(p_0,\xi_0) \frac{\partial P(p_0,\xi_0,\tau,p_{RE})}{\partial p_{RE}} d^3p_0
    . \label{Adjoint energy distribution}
\end{eqnarray}
Here, the derivative of the RPF is computed by automatic differentiation, with $p_{RE}$ now being an input to the neural network. An additional advantage of this approach is that the high energy boundary terms appearing in Eq. (\ref{RunawayAdjointRelationGeneral2}) are not present. Specifically, by passing the definition of the adjoint in Eq. (\ref{RunawayAdjointRelationGeneral2}) into Eq. (\ref{Adjoint energy distribution}), the high energy boundary 
term vanishes and the Heaviside function in the terminal condition converts to a Dirac-delta term.
Taking this limit simplifies the energy integral so that Eq. (\ref{Adjoint energy distribution}) is equivalent to a pitch-angle-averaged Green's function.
The vanishing high energy boundary term implies that if this process is applied to any moment $M$, the boundary term previously discussed is no longer present.
The treatment of the high-energy boundary condition should not change, however, so that a single model can predict both the density moment for a fixed value of $p_{RE}$ and the energy distribution across variable $p_{RE}$.

\section{Physics-Informed Neural Networks}
\label{sec:PINNs}
Physics-informed neural networks have emerged as a prominent example of physics-constrained deep learning methods~\cite{raissi_perdikaris_karniadakis_2019,Karniadakis_2021, lagaris_likas_fotiadis_1998,karpatne_atluri_faghmous_steinbach_banerjee_ganguly_shekhar_samatova_kumar_2017,lusch_kutz_brunton_2018,wang_kashinath_mustafa_albert_yu_2020}.
The aim of the present section is to extend the generalizability of the adjoint framework by utilizing a PINN to capture moments and the energy distribution of runaways by 
solving the adjoint solutions described in Section \ref{sec:adjointFP} across a wide range of plasma parameters including electric field strength $E_{\Vert}$, effective charge state $Z_{eff}$, synchrotron strength $\alpha$, and for the case where we would like to infer the energy distribution, $p_{RE}$.
Once trained, the resulting framework will allow for nearly
instantaneous projections of the discussed RE quantities incorporating fully kinetic physics.

\subsection{Physics-Constrained Deep Learning}
\label{subsec:physics_constrained}
The present discussion will only focus on
the essential concepts, where the interested reader is referred to Ref. \cite{Karniadakis_2021} 
for a more detailed discussion.
Here, the underlying strategy is to use physical constraints (primarily PDEs) to 
regularize the training of a neural network. 
Such an approach not only provides a natural means of avoiding overfitting, thus providing a more robust 
interpolative tool, but also opens up the possibility of greater generalizability to unseen parameter regimes. A
PINN in its simplest form can be expressed as
\begin{align}
    Loss= \frac{1}{N_{PDE}}\sum_i^{N_{PDE}} \mathcal{R}^2 \left(\mathbf{p}_i, \bm{\lambda}_i,\tau_i \right) + \frac{1}{N_{IC}}\sum_i^{N_{IC}}\left[G_i-G(\mathbf{p}_i, \bm{\lambda}_i, \tau_i=0)\right]^2 + \frac{1}{N_{BC}}\sum_i^{N_{BC}}\left[G_i-G(\mathbf{p}_i, \bm{\lambda}_i,\tau_i)\right]^2,
    \label{eq:deeplearning_loss}
\end{align}
where the first term on the RHS represents points inside the domain minimizing the residual of the adjoint PDE in Eq. (\ref{eq:adjointFP1}), and $G_i$ represents data points enforcing initial and boundary conditions.
We note that our adjoint framework works in reverse time where the initial condition is changed to a 
terminal condition at $t_{final}$, such that we use the temporal variable $\tau= t_{final}-t$, and the second loss term on the RHS in Eq. (\ref{eq:deeplearning_loss}) satisfies the terminal condition discussed in Sec. \ref{sec:adjoint_solution}.
The quantity $\bm{\lambda}$ represents plasma parameters which will be taken to be $\bm{\lambda}=\left(E_\Vert,Z_{eff},\alpha\right)$, or $\bm{\lambda}=\left(E_\Vert,Z_{eff}, \alpha, p_{RE} \right)$ in Sec. \ref{sec:energy_distrib} where the energy distribution is inferred,
$N_{IC}$ is the number of initial condition points, $N_{BC}$ is the number of boundary condition points,
$N_{PDE}$ is the number of domain points. 
We note that the model is trained in the absence of data and thus relies solely on physics constraints.
Additional constraints will be added (see Sec. \ref{sec:PL} below) to enhance the training of the model and prevent predictions beyond the expected range.
This prevents the PINN from making unphysical predictions while simultaneously severely restricting the solutions that the optimizer searches across, leading to more robust convergence of the PINN. 

\subsection{\label{sec:PL}Physics Layers}
\label{subsec:physics_layer}
As a means of improving the accuracy and robustness of the PINN, we implement several properties of the solution as hard constraints via the 
addition of a ``physics layer'' to the neural network (NN). This additional layer takes the output of the hidden layers of the NN and imposes a series of 
transforms that normalize the solution to an expected range, ensuring that the lower-momentum boundary conditions are exactly satisfied. For all adjoint problems we will include a transform of the form:
\begin{equation}
    G \left(p,\xi,\tau;\mathbf{\lambda} \right) = \tanh \left[ \left(\frac{p-p_{\min}}{p_{\max}-p_{\min}} \right)G_{NN}^n \right],
\end{equation}
where $G_{NN}$ is the output of the hidden layers of the neural network, 
$n$ is designed to be either $n=1$ to constrain the solution between negative one and one, or $n=2$ to constrain the solution between zero and one.
The output transform here is designed to automatically satisfy the lower momentum boundary condition in which $G(p=p_{\min})=0$.
The terminal condition is enforced through loss terms where points are sampled on the terminal condition 
boundary and are enforced to satisfy 
\begin{equation}
    G(\tau=0)= g \left(p,\xi \right)\frac{1}{2}\left[1-\tanh \left(N\frac{p-p_{RE}}{p_{\max}-p_{\min}} \right)\right].
    \label{eq:terminal_cond2}
\end{equation}
Here, $N$ controls how sharp the transition is; it is taken to be large ($N>>1$) and tuned for each physical quantity.
We choose to enforce the high energy boundary condition as a soft constraint in the loss function due to the narrow width of the region of interest for 
particles flowing through the high energy boundary.
We treat this narrow region by sampling points directly on the positive boundary discussed in Eq. (\ref{Up_equation}).
An additional layer is added before the hidden layers and after the inputs such that an additional coordinate $g(p,\xi)$ is added 
to improve the network convergence,
\begin{equation}
    \left[p,\xi,\tau,\bm{\lambda} \right] \rightarrow \left[p,\xi,\tau,\bm{\lambda}, g \left(p,\xi \right) \right].
    \label{input_transform}
\end{equation}
Here, $g(p,\xi)$ is the weighting factor for each moment discussed in previous sections.
The range of energy ($p$) in the PINN is chosen to be between $10$ keV and $16$ MeV and the time range is chosen to be from $\tau=0$ to $\tau= 20 \tau_c$. 
We chose to train with the SOAP optimizer using the pytorch-optimizer library \cite{SOAP} where the model architecture consists of a relatively large network of 64 neurons per layer and 6 layers.
Training was performed with five million PDE training points sampled via a Sobol distribution with
an additional 250 thousand points added to the $\xi=-1, \xi=1$ boundaries, terminal condition and high energy boundary condition ($p_{\max}$ and $U_p>0$) respectively.
For calculating moments we choose $p_{RE}\approx 2.78$ or 1 MeV where the vector $\bm{\lambda}$ in Eq. (\ref{input_transform}) contains the plasma parameters $(E_{\Vert},Z_{eff},\alpha)$ where $E_{\Vert}$ ranges from $1$ to $4$, $Z_{eff}$ ranges from $1$ to $5$, and $\alpha$ ranges from $0.05$ to $0.2$.

\subsection{Adaptive Sampling}
As a means of focusing on regions of interest in the PINN training, we implement an adaptive resampling algorithm inspired by \cite{wu_zhu_tan_yadhu_kartha_lu_2022}. 
Training points are sampled via a quasirandom Sobol distribution, the network trains for a fixed number of epochs, followed by 
an adaptive resampling of training points informed by the loss in Eq. (\ref{eq:deeplearning_loss}),
\begin{equation}
    L_{adaptive}(\mathbf{x}) = \left[ \frac{L(\mathbf{x})}{<|L(\mathbf{x})|>} \right]^k + c,
\end{equation}
where $L_{adaptive}$ is an unnormalized probability distribution to resample points, $L(\mathbf{x})$ is the residual loss in Eq. (\ref{eq:deeplearning_loss}),
and $<|L(\mathbf{x})|>$ is the mean of the absolute value of the residual. 
The $k$ parameter skews towards sampling points in high residual regions and is chosen to be $k=2$, the $c$ parameter adds a uniform probability everywhere on the domain and is chosen to be $c=1$.

\subsection{JONTA: a particle-based runaway solver}
\label{sec:JONTA}
To verify our PINN implementation, we compare the derived results against a JAX-based kinetic solver optimized to run on GPU accelerated hardware initially described in the context of fast ion transport in Ref. \cite{mcdevitt_arnaud_2026} in which 
the deterministic portion is solved via a four-stage Runge--Kutta (RK4) integration scheme, 
\begin{eqnarray}
    \frac{d\gamma}{dt} = \frac{p}{\gamma} \left[-E_\Vert \xi -\frac{1+p^2}{p^2} -\alpha p \gamma (1-\xi^2) \right],\\
    \frac{d \xi}{dt} = -\frac{E_\Vert}{p}(1-\xi^2) + \alpha \xi \frac{(1-\xi^2)}{\gamma},
\end{eqnarray}
and the Lorentz collision operator is implemented by a Monte Carlo equivalent~\cite{Boozer:1981},
\begin{equation}
    \xi_{n+1} = \xi_{n}(1-\nu_D\Delta t)\pm \sqrt{(1-\xi_n^2)\nu_D\Delta t},
\end{equation}
where $\Delta t$ is the collisional time step, taken to be $\Delta t= 10^{-3} \tau_c$.
We vary a wide range of initial conditions and plasma parameters to demonstrate the robustness of the PINN model.
While the pitch initialization varies, the initial energy distribution is chosen to be a Gaussian centered at 8 MeV or $p' \approx 16.6$ with a standard deviation of $\sigma_p = 3.5$, i.e.
\begin{equation}
    f_{init}(p_0,\xi_0)=\frac{1}{\sqrt{2\pi \sigma_p}}e^{-\left(\frac{p_0-p'}{\sigma_p}\right)^2}f(\xi_0),
    \label{eq:finit_init_cond}
\end{equation}
where we demonstrate the applicability by changing pitch dependent function $f(\xi_0)$ to be arbitrary along the $\xi$ axis.
\begin{figure}[!htbp]
    \subfigure[]{\includegraphics[scale=0.33]{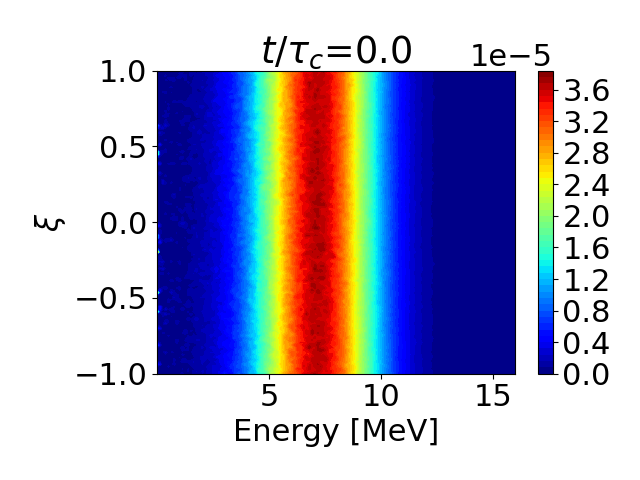}}
    \subfigure[]{\includegraphics[scale=0.33]{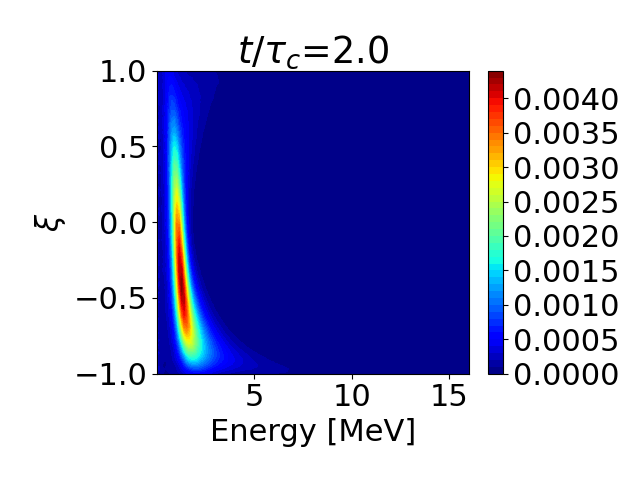}}
    \subfigure[]{\includegraphics[scale=0.33]{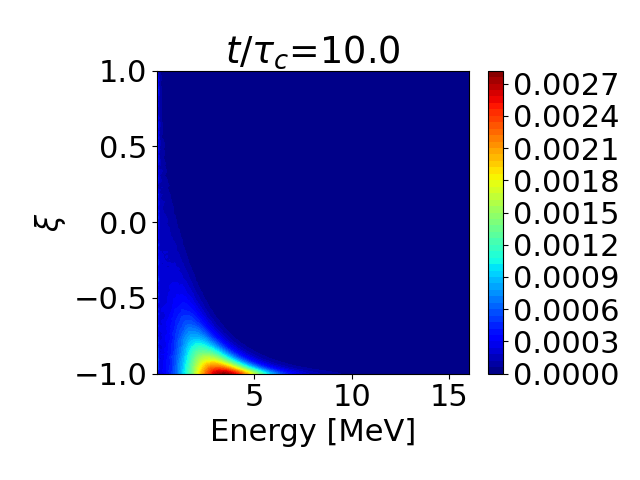}}
    \caption{Example of a JONTA RE Monte Carlo simulation where $E_\Vert = 2.5$, $Z_{eff}=3.0$, $\alpha=0.1$. Isotropic initial condition described in Eq. (\ref{eq:finit_init_cond}) (Panel a.), particle distribution at time $t/\tau_c=2$ (Panel b.), particle distribution at time $t/\tau_c=10$ (Panel c.)}
    \label{fig:JONTA_example}
\end{figure}

The adjoint formulation for runaways described in Sec. (\ref{sec:adjoint_solution}), can be interpreted directly by summing the weights of marker particles by their respective phase space coordinate,
\begin{equation}
    M(t) = \frac{1}{N}\sum_{i=0}^{N} g(p_i,\xi_i,t_i)\Theta(p_i-p_{RE}),
    \label{eq:JONTA_moments}
\end{equation}
where $M(t)$ is the moment first described in Eq. (\ref{eq:Moment_full_distribution}).
We verify the adjoint-PINN method by interpreting Eq. (\ref{eq:TAS3}) such that the first term on the RHS counts particles above $p_{RE}$ and the second term refers to freezing particles once they cross the high energy boundary where calculating the Monte Carlo moments requires applying these interpretations.
Likewise, we compare moment predictions in which particles are \emph{not} frozen when they exit the high energy boundary and later show that if $g(p,\xi)$ doesn't significantly vary in energy, then the adjoint method agrees well with Monte Carlo; however, if $g(p,\xi)$ indeed varies in energy, then the adjoint method can result in poor agreement compared to not freezing particles and is due to significant contributions when particles exit the boundary.
Pushing particles, binning, and generating plots with JONTA takes roughly 3.5 minutes on a B200 GPU with ten million particles, whereas evaluating a converged PINN takes milliseconds. Training a parametric PINN takes roughly 14 hours on the same B200 GPU. 


\section{\label{sec:REM}Runaway Electron Moments}
\label{sec:moments_results}

\subsection{Current Moment}
\label{subsec:current_moment}
The current probability function (CPF) is trained to capture the RE parallel current moment by treating the terminal condition as a Heaviside function centered at $p_{RE}$ 
which returns 0 for $p<p_{RE}$ and $-v_\Vert$ for $p>p_{RE}$ where $v_\Vert=v\xi$, as described in Eq. (\ref{eq:TAS6}), i.e.
\begin{equation}
    C(\tau=0)= \frac{-v\xi}{2}\left[1-\tanh \left(N\frac{p_{RE}-p}{p_{\max}-p_{\min}}\right)\right],
\end{equation}
where $N = 80$.
The low-energy boundary condition is enforced to be $C(p_{\min},\xi)=0$ implying that electrons which decelerate through $p_{\min}$ do not contribute to the RE current. 
An output transform is chosen to keep the CPF solution bounded between $-1$ and $1$ and to act as a hard-constraint for the low energy boundary condition,
\begin{equation}
    C = \tanh\left[\frac{p-p_{\min}}{p_{\max}-p_{\min}}C_{NN}\right],
\end{equation}
where $C_{NN}$ is the output of the hidden layers.
The high-energy boundary condition is enforced as a soft constraint in the loss function to be $C(p_{\max})=-v_\Vert$ in regions with $U_{p}>0$, and is otherwise left unconstrained.
Once electrons pass through this boundary their velocity and pitch distribution is fixed. This is reasonable when $p_{\max}$ is sufficiently large such that electrons have $\left| v_\Vert \right| \approx c$ at the high energy boundary.
Adaptive sampling was performed every 1000 epochs for 134 thousand epochs where the PDE test loss reached $1.01\times 10^{-7}$, the terminal condition test loss reached $1.74\times 10^{-8}$,
and the BC test loss is $9.74\times 10^{-8}$.

\begin{figure}[!htbp]
    \subfigure[]{\includegraphics[scale=0.33]{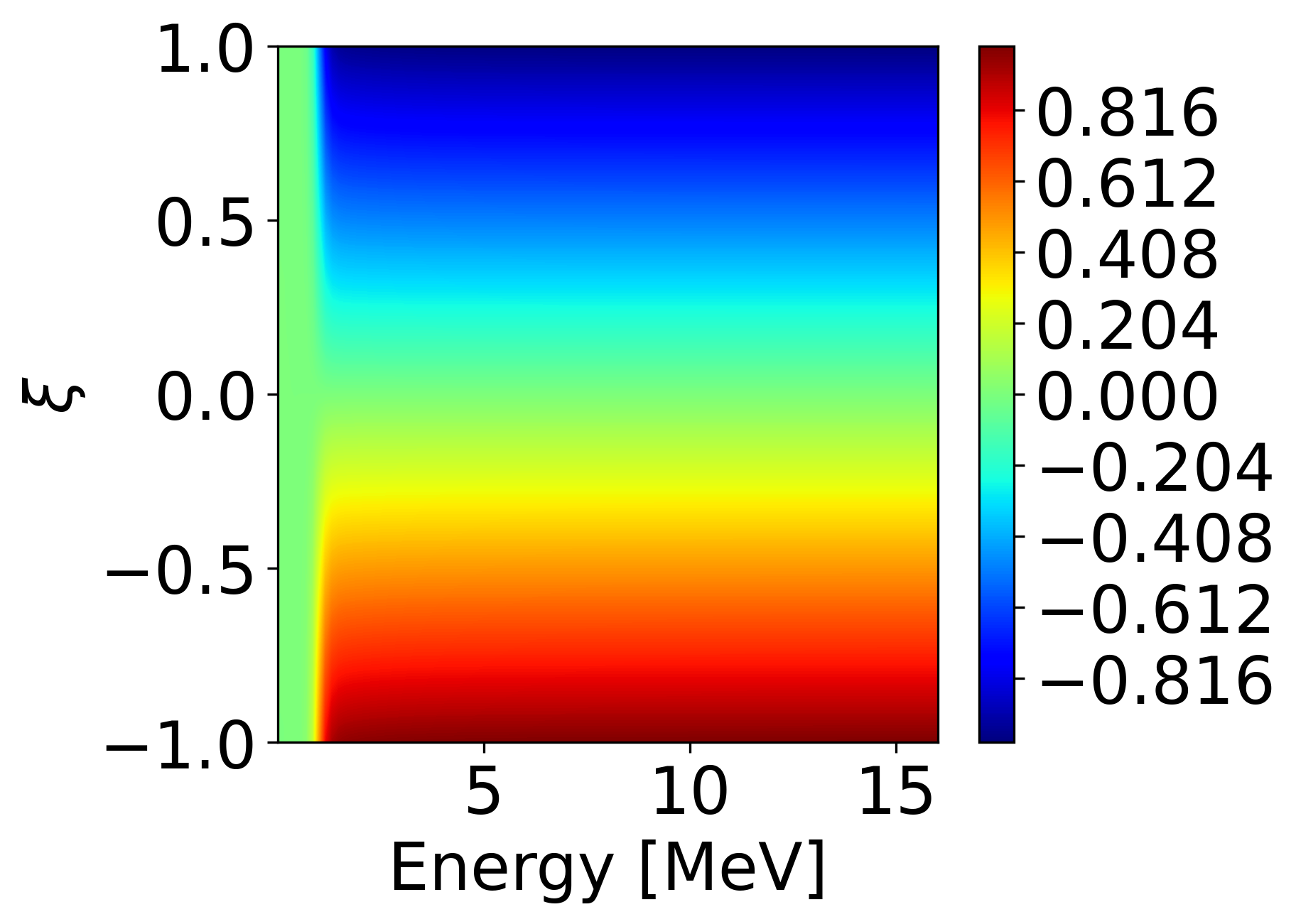}}
    \subfigure[]{\includegraphics[scale=0.33]{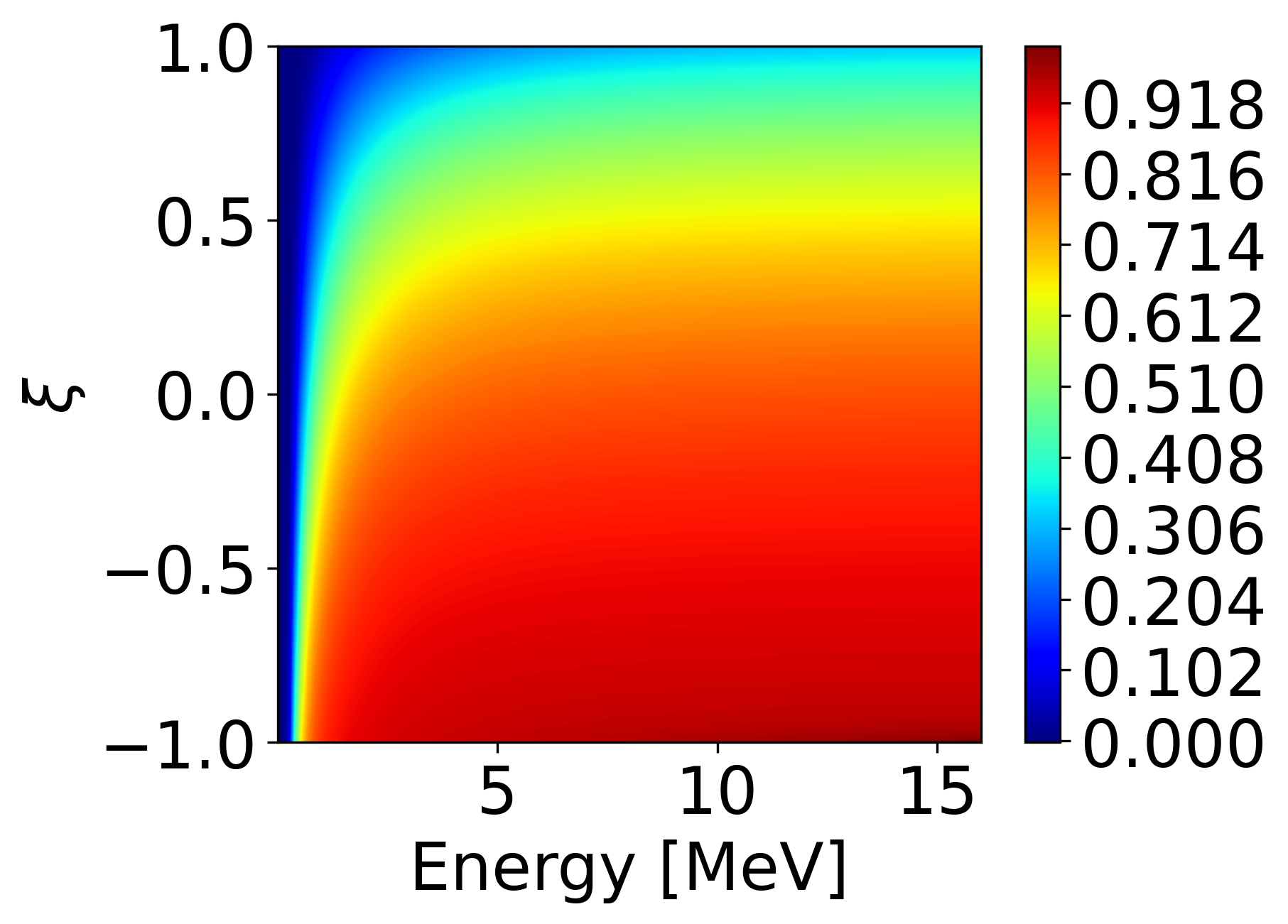}}
    \subfigure[]{\includegraphics[scale=0.30]{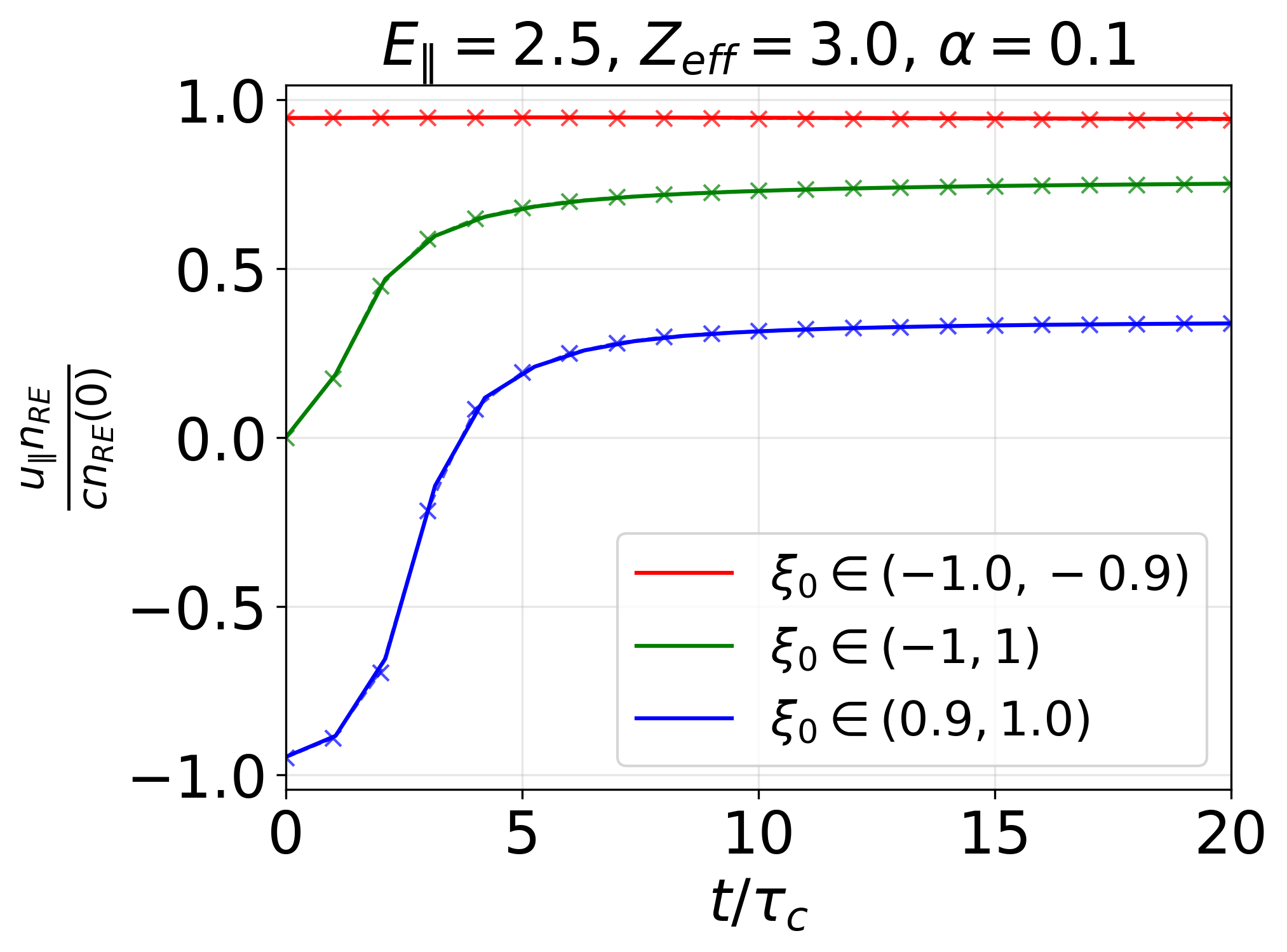}}
    \subfigure[]{\includegraphics[scale=0.37]{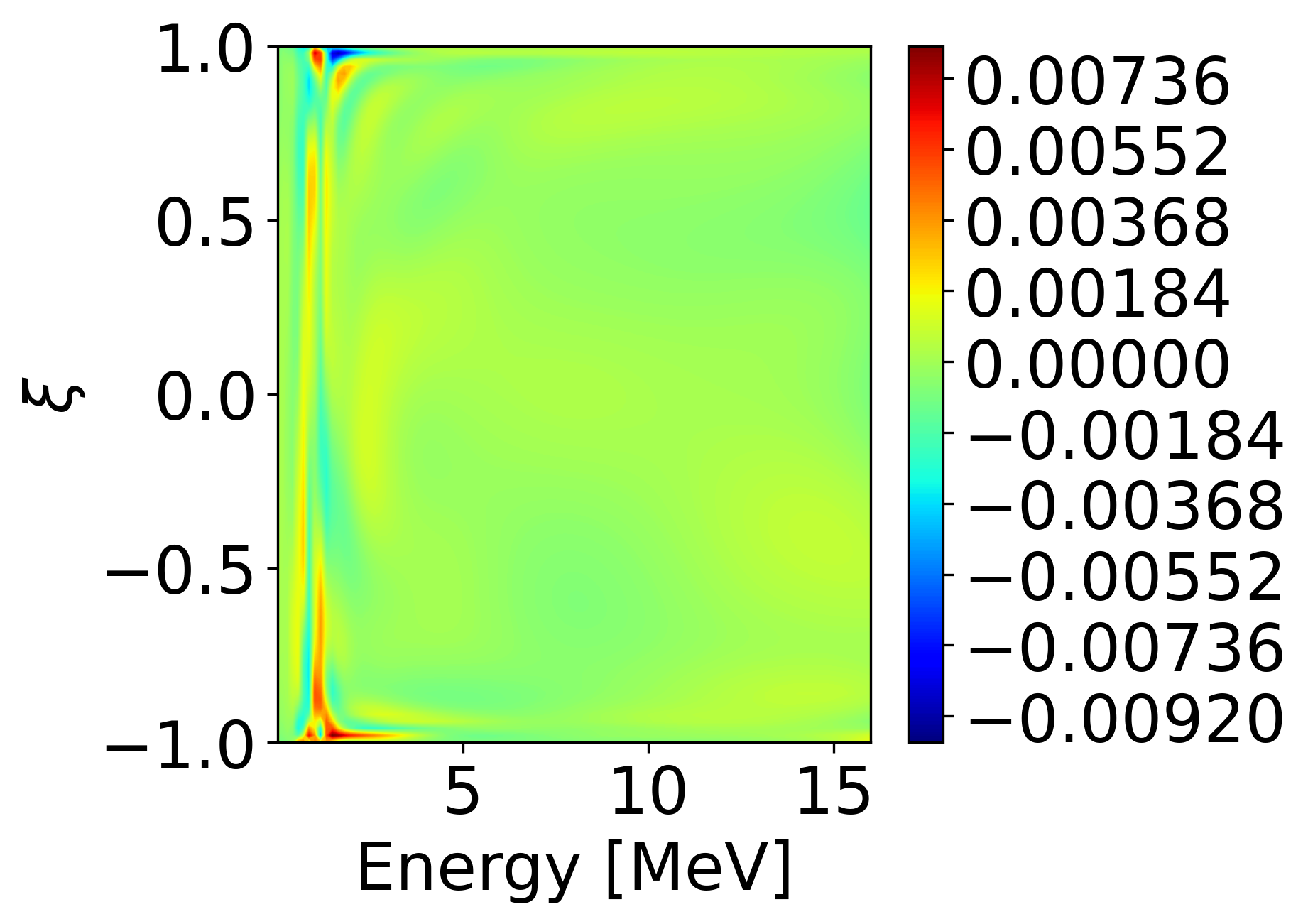}}
    \subfigure[]{\includegraphics[scale=0.37]{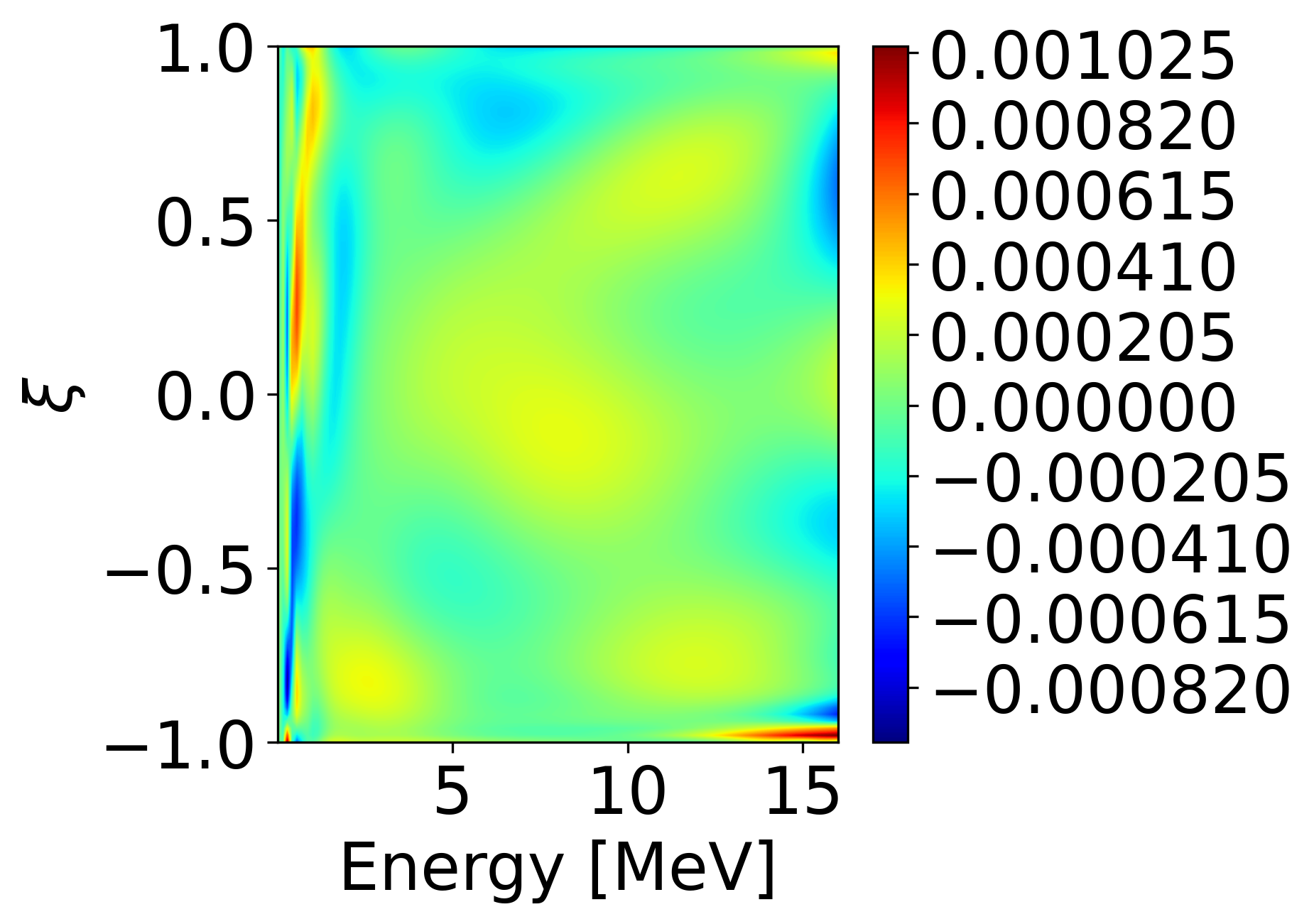}}
    \subfigure[]{\includegraphics[scale=0.30]{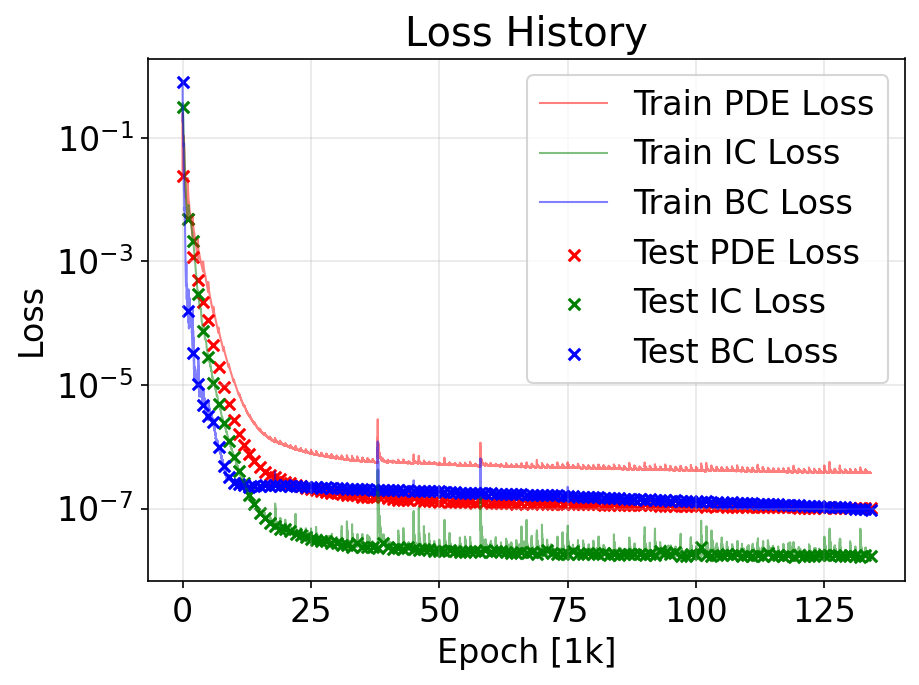}}
    \caption{Panels (a--e) evaluate the CPF for $E_\Vert = 2.5$, $Z_{eff}=3$, $\alpha=0.1$.
    Panels (a) and (d) show the CPF and residual evaluated at the terminal condition; panels (b) and (e) show the CPF evaluated at $\tau/\tau_c = 20$; 
    panel (c) shows the predicted runaway current for various initial pitch orientations $f(\xi_0)$ and an energy distribution as discussed in Sec.~\ref{sec:JONTA}---PINN [solid], Monte Carlo with frozen particles [dashed], and Monte Carlo without frozen particles [scatter].
    Panel (f) shows the loss history of the PINN.} 
    \label{velocity_moment_fig1}
\end{figure}

From Fig.\ \ref{velocity_moment_fig1}(c), each curve corresponds to a different initial pitch distribution $f(\xi_0)$ with the initial energy distribution discussed in Sec.~\ref{sec:JONTA}. Specifically, the red curve corresponds to a uniform $f(\xi_0)$ initialization distributed between $-1$ and $-0.9$, where the runaway current remains nearly constant, the green curve corresponds to an isotropic initialization for which the average current is initially zero but later turns positive, and the blue curve corresponds to a uniform initialization distributed between $0.9$ and $1$ and corresponds to a negative initial current.
Excellent agreement between the predictions of the runaway current PINN and JONTA simulations are evident in Fig. \ref{velocity_moment_fig1}(c).
Here,  
the adjoint-PINN method is plotted as a solid line and the Monte Carlo points are plotted as `x' markers. This good agreement indicates that the high energy boundary is sufficiently high such that most runaways remain inside the domain as discussed in Sec. \ref{sec:TAS}.


\begin{figure}[!htbp]
    \subfigure[]{\includegraphics[scale=0.32]{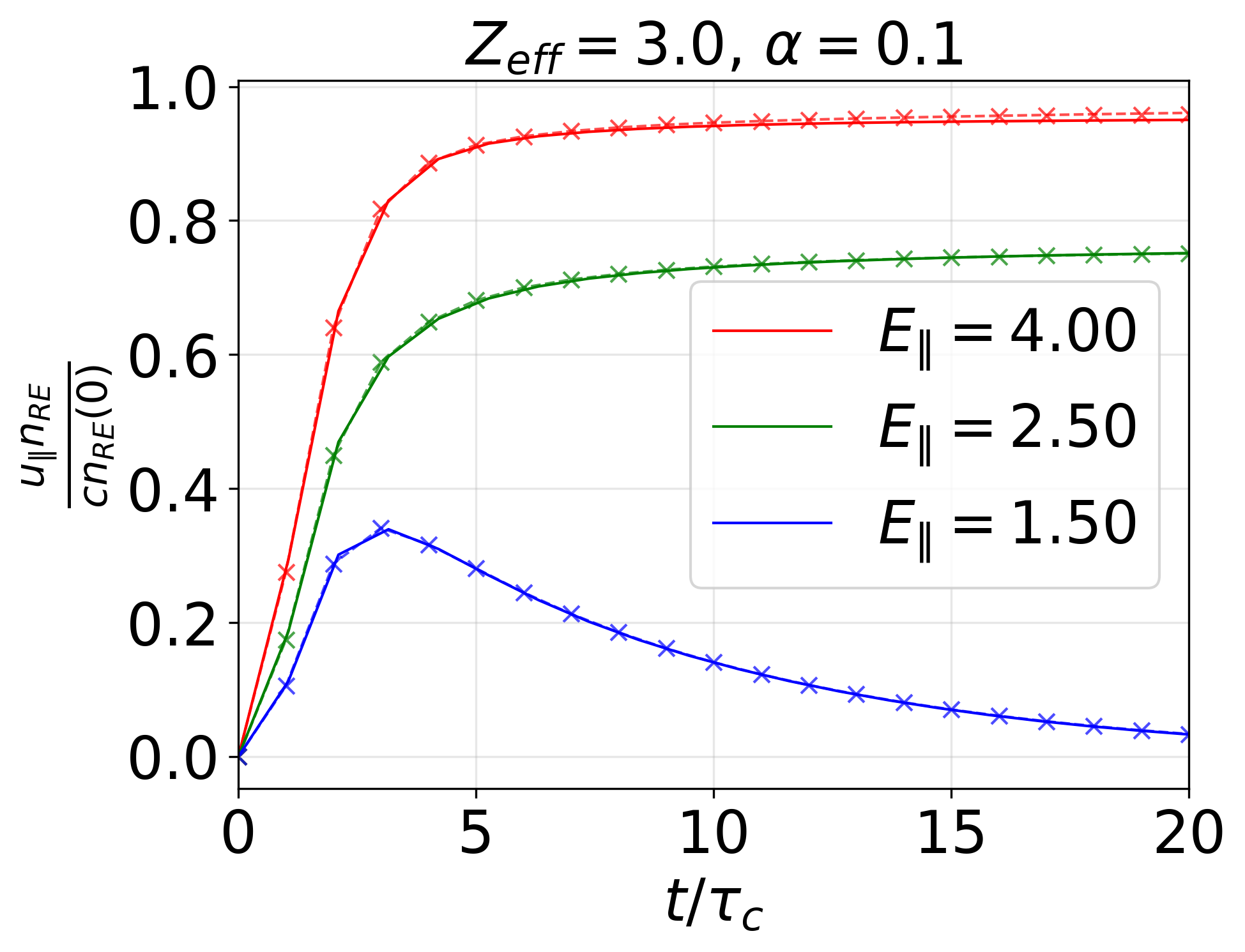}}
    \subfigure[]{\includegraphics[scale=0.32]{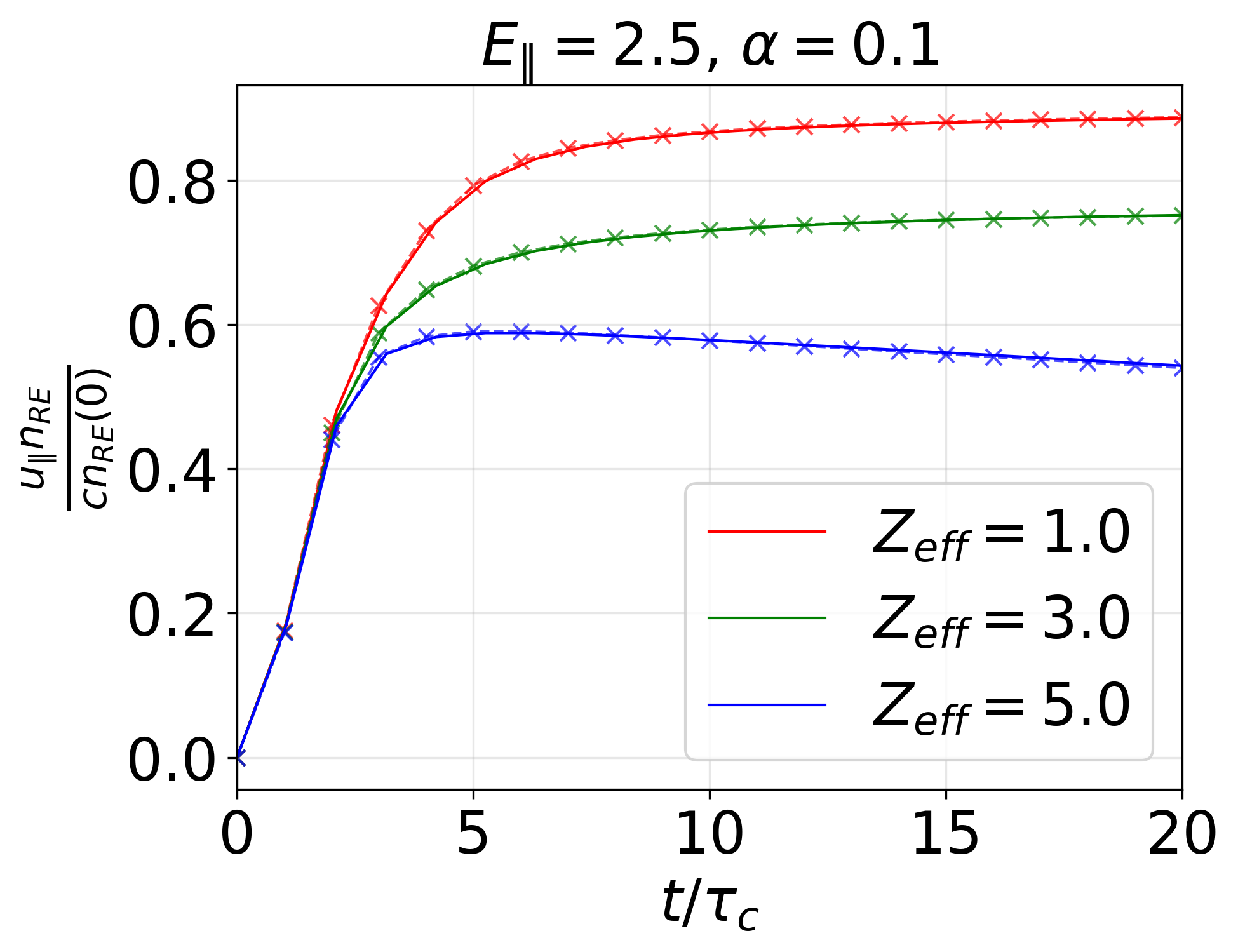}}
    \subfigure[]{\includegraphics[scale=0.32]{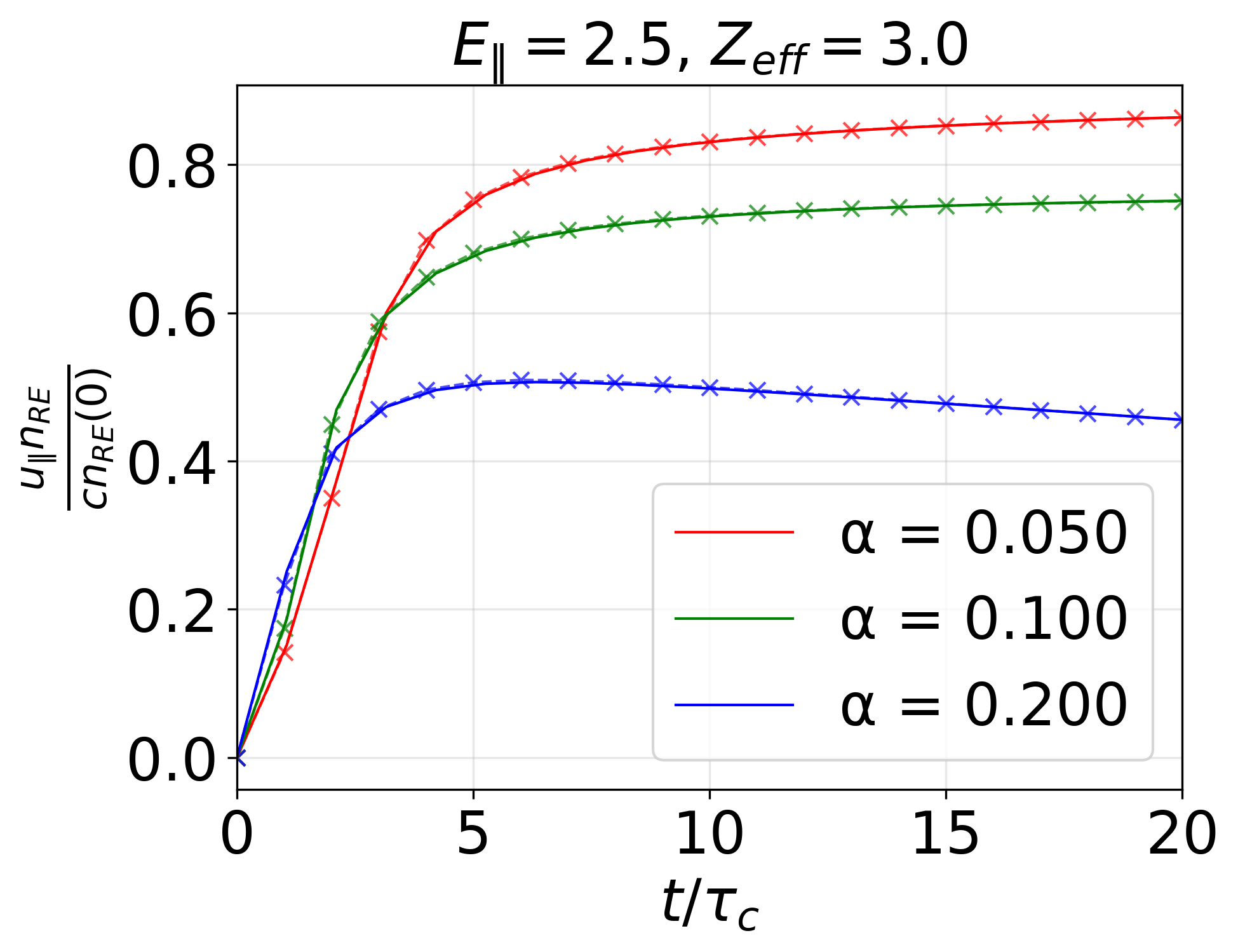}}
    \caption{Runaway current versus time for variations in $E_\Vert$ (a), $Z_{eff}$ (b), and $\alpha$ (c) for isotropic initialization and an energy distribution as discussed in Sec.~\ref{sec:JONTA}---PINN [solid], Monte Carlo with frozen particles [dashed], Monte Carlo without frozen particles [scatter].} 
    \label{current_moment_results}
\end{figure}

Figure \ref{current_moment_results} illustrates the time evolution of runaways as the electric field $E_\Vert$, effective charge $Z_{eff}$, and strength of synchrotron radiation $\alpha$ are varied, allowing for the threshold electric field to be identified.
For the parameters plotted in Fig.\ \ref{current_moment_results}(a), the threshold electric field is roughly $E_{\Vert,\mathrm{thresh}} = 1.84$; particles remain runaways above this threshold and the current saturates, whereas for $E_\Vert=1.5$, the current decays as particles collisionally slow into the bulk.
The combined results of Figs. \ref{current_moment_results} and \ref{velocity_moment_fig1} (c) show that the adjoint-PINN framework can accurately characterize runaway current for arbitrary initial conditions as well as a wide range of plasma parameters in a single model.

\FloatBarrier
\subsection{Energy Moment}
\label{subsec:energy_moment}
This energy probability function (EPF) is trained to capture the average kinetic energy of runaways by treating the terminal condition 
as a Heaviside function such that the solution returns $0$ for $p<p_{RE}$ and $m_ec^2(\gamma-1)$ for $p>p_{RE}$, i.e.
\begin{equation}
    \mathcal{E}(\tau=0)=m_ec^2 \left(\gamma-1 \right)\frac{1}{2}\left[1-\tanh{\left(N\frac{p_{RE}-p}{p_{\max}-p_{\min}}\right)}\right],
\end{equation}
where we choose $N=32$. 
An additional input variable was added corresponding to the energy coordinate $m_ec^2\left(\gamma-1\right)$ as discussed in \ref{subsec:physics_layer} and an output transform is designed to enforce that the solution is positive and that the $p_{\min}$ boundary condition is automatically enforced,
\begin{equation}
    \mathcal{E} = \left(\frac{p-p_{\min}}{p_{\max}-p_{\min}}\right) \mathcal{E}_{NN}^2,
\end{equation}
where $\mathcal{E}_{NN}$ are the outputs of the neural network and $\mathcal{E}$ is the resulting solution.
The high energy boundary condition $\mathcal{E}(p_{\max},\tau) = m_ec^2(\gamma-1)$ when $U_p>0$ is enforced via a loss term which assumes that electrons that exit the domain at the energy of that boundary (e.g.\ for $p_{\mathrm{\max}}$ corresponding to 16 MeV) are assigned 16 MeV thereafter in the surrogate.
Adaptive sampling was performed every 2,500 epochs for 
180,000 epochs with a test PDE loss of $1.99\times 10^{-5}$, a terminal condition test loss of $6.24\times 10^{-6}$ and 
a boundary condition test loss of $1.32\times 10^{-5}$. 

\begin{figure}[!htbp]
    \subfigure[]{\includegraphics[scale=0.33]{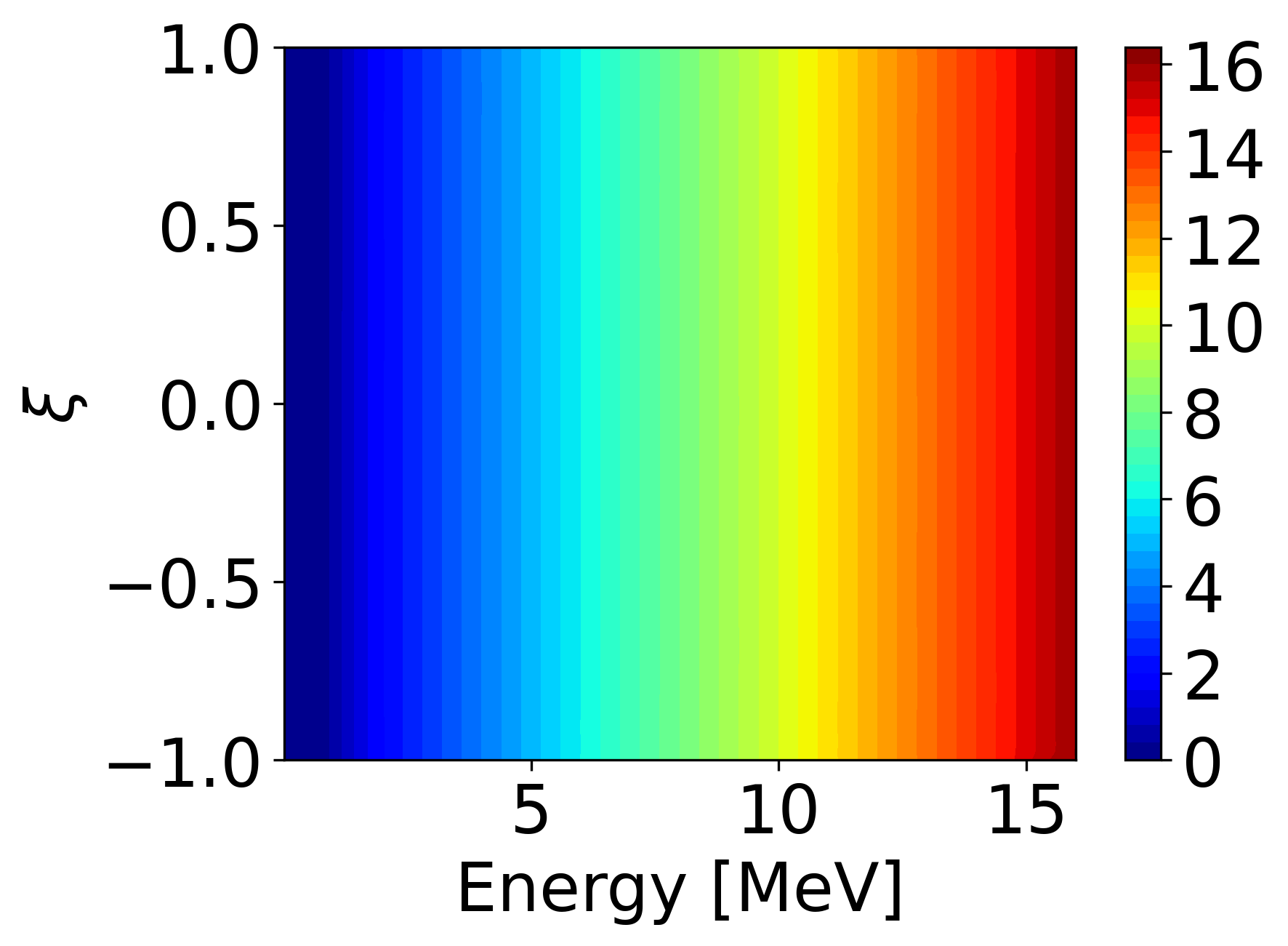}}
    \subfigure[]{\includegraphics[scale=0.33]{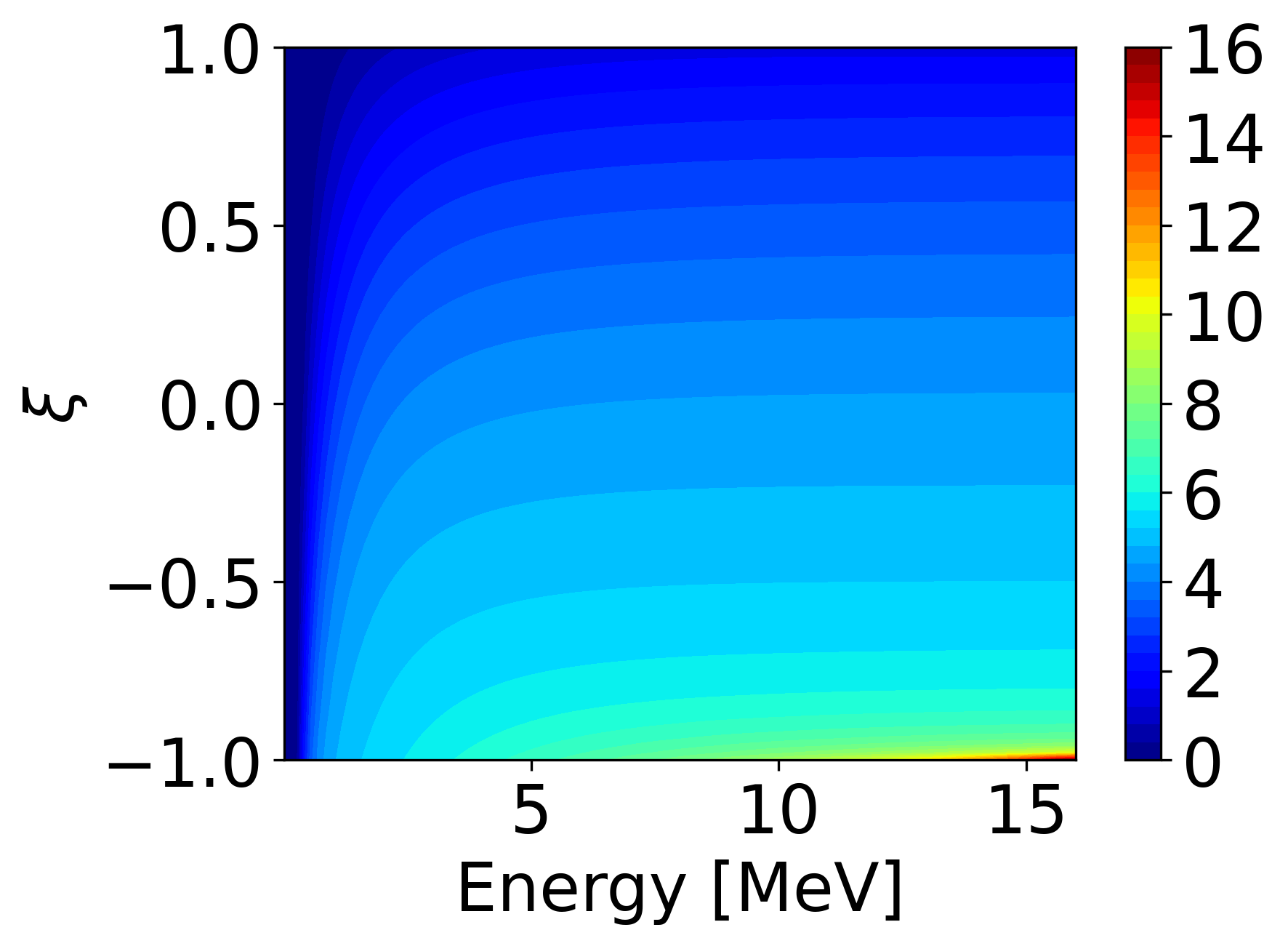}}
    \subfigure[]{\includegraphics[scale=0.30]{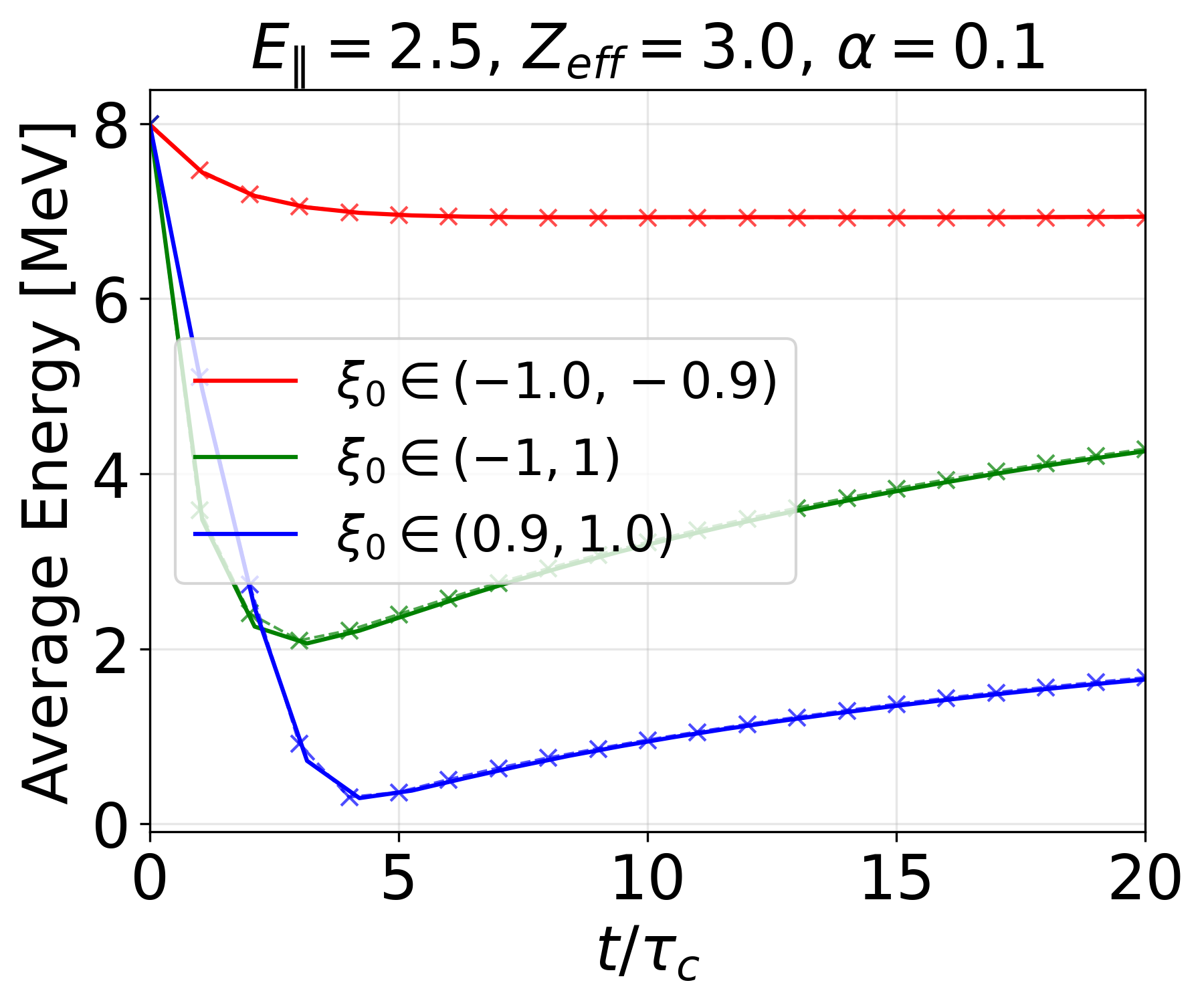}}
    \subfigure[]{\includegraphics[scale=0.37]{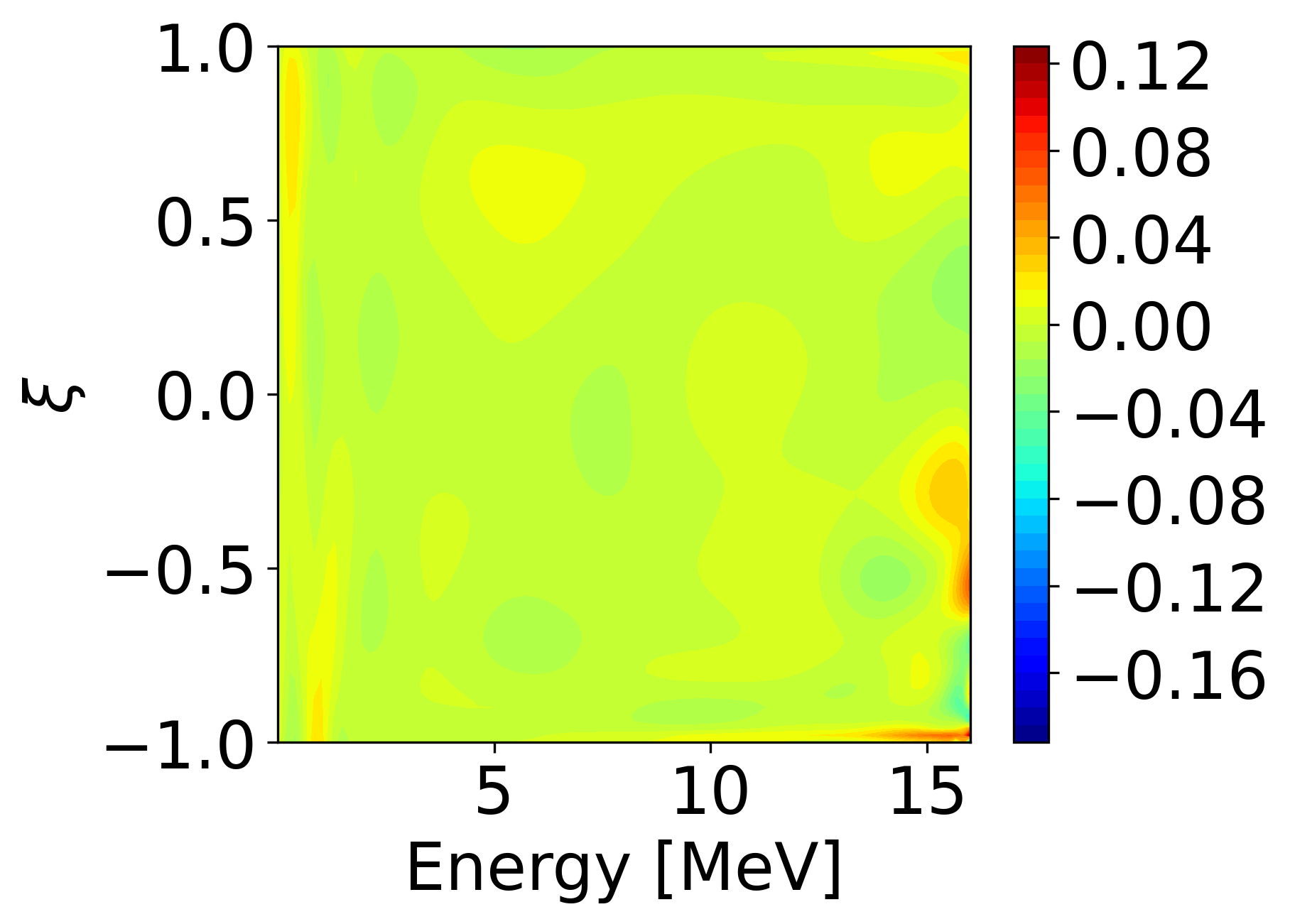}}
    \subfigure[]{\includegraphics[scale=0.37]{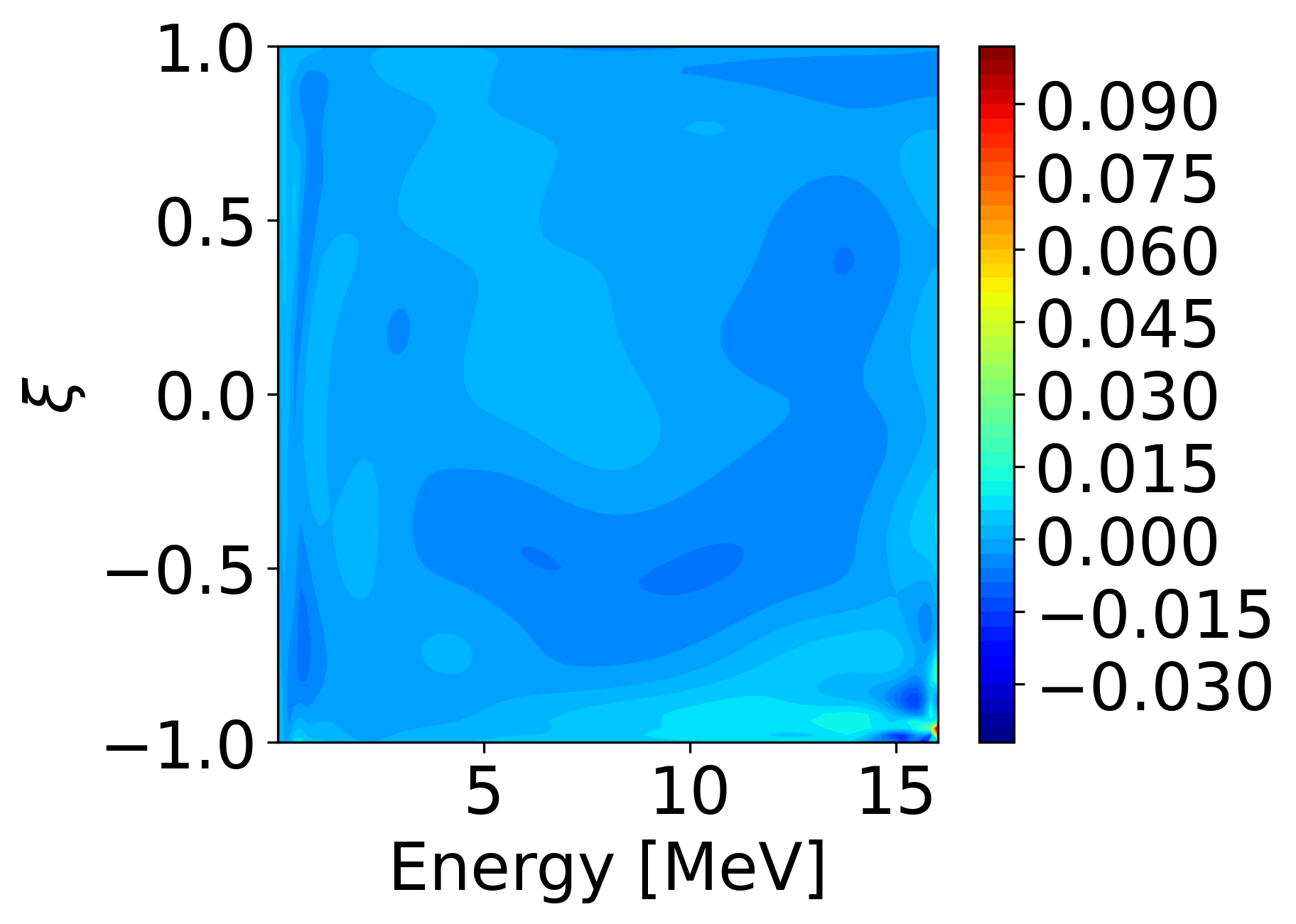}}
    \subfigure[]{\includegraphics[scale=0.30]{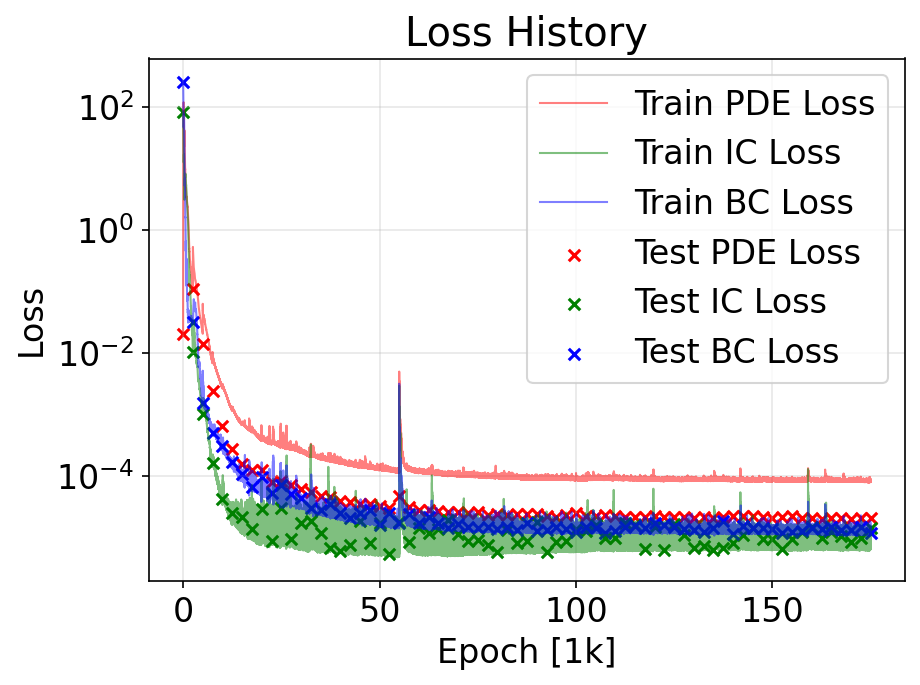}}
    \caption{Panels (a--e) evaluate the EPF for $E_\Vert = 2.5$, $Z_{eff}=3$, $\alpha=0.1$.
    Panels (a) and (d) show the EPF and residual at the terminal condition ($\tau/\tau_c=0$); panels (b) and (e) show them at ($\tau/\tau_c=20$).
    Panel (c) shows the evolution of the average energy for various initial pitch orientations $f(\xi_0)$ and an energy distribution as discussed in Sec.~\ref{sec:JONTA}---PINN [solid], Monte Carlo with frozen particles [dashed], and Monte Carlo without frozen particles [scatter].
    Panel (f) shows the PINN loss history.}
    \label{energy_moment_fig}
\end{figure}
Figure~\ref{energy_moment_fig} (c) demonstrates that the adjoint-PINN method can accurately capture the average energy across arbitrary initial conditions
whereby the average energy of runaways above $1$ MeV is plotted for an initial energy distribution discussed in Sec.~\ref{sec:JONTA} 
and for arbitrary initial pitch distributions discussed with the current prediction in the previous section. 
From Fig.~\ref{energy_moment_fig} (f), we note that the final loss values of the energy moment are higher than the current moment due to an order of magnitude difference in the solution range, which when passed through a mean-squared error loss, produces two orders of magnitude difference.

\begin{figure}[!htbp]
    \subfigure[]{\includegraphics[scale=0.33]{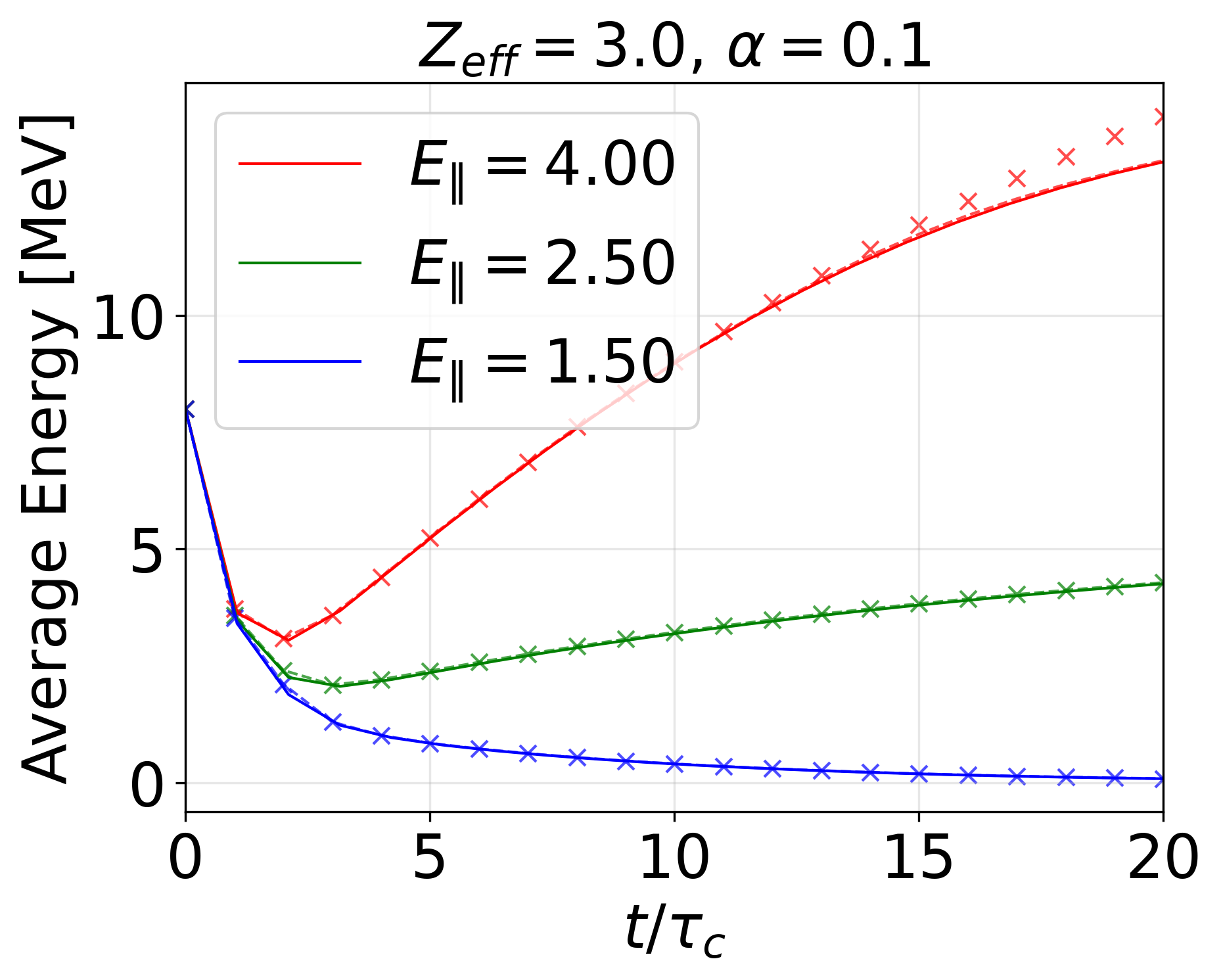}}
    \subfigure[]{\includegraphics[scale=0.33]{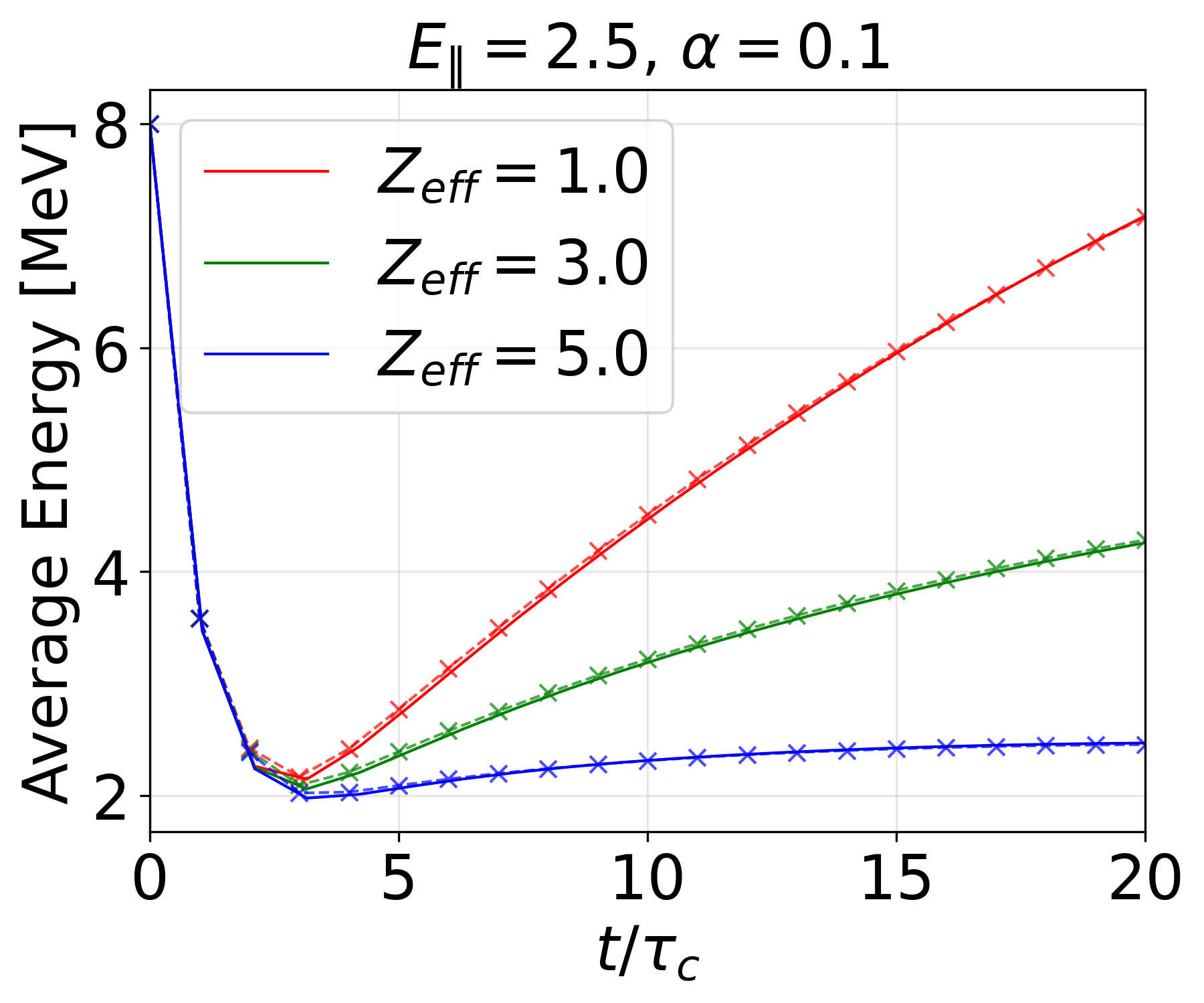}}
    \subfigure[]{\includegraphics[scale=0.33]{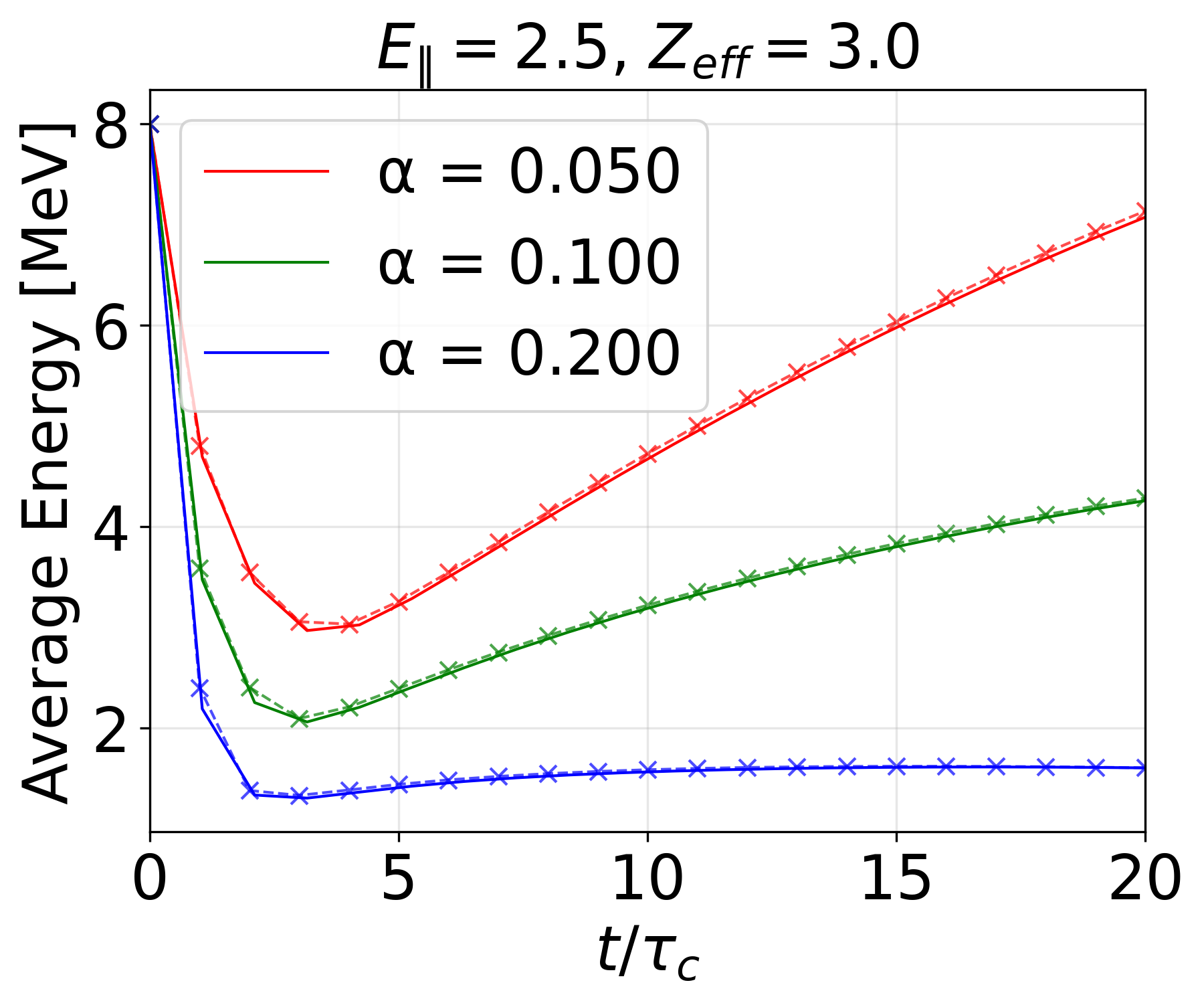}}
    \caption{Runaway average energy versus time for variations in $E_\Vert$ (a), $Z_{eff}$ (b), and $\alpha$ (c) under isotropic initialization and an energy distribution as discussed in Sec.~\ref{sec:JONTA}: PINN [solid], Monte Carlo with frozen particles [dashed], Monte Carlo without frozen particles [scatter].}
    \label{energy_moment_fig_variation}
\end{figure}

For modest plasma parameters that keep runaways inside the computational domain, the results in Fig.~\ref{energy_moment_fig_variation}(b,c) show excellent agreement with Monte Carlo scatter points and indicates that the adjoint-moment PINN can robustly serve as a rapid surrogate for average runaway energy.
However, when runaways exit the domain, the adjoint surrogate assigns them $16$ MeV, which leads to poor agreement for strong electric fields or weak synchrotron damping, as shown in Fig.\ \ref{energy_moment_fig_variation}(a). 

\FloatBarrier
\section{Runaway Electron Energy Distribution}
\label{sec:energy_distrib}
\begin{figure}
    \subfigure[]{\includegraphics[scale=0.33]{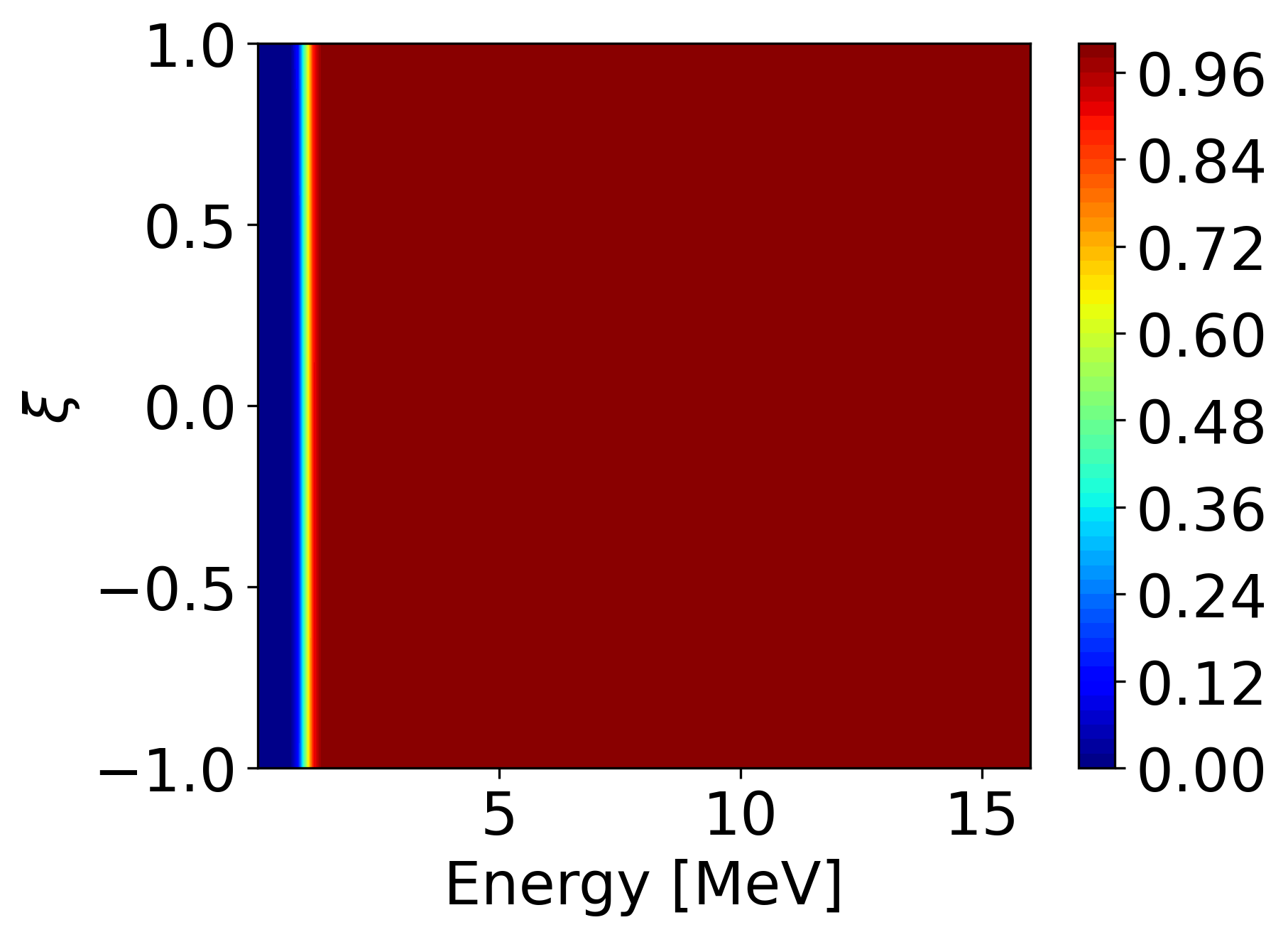}}
    \subfigure[]{\includegraphics[scale=0.33]{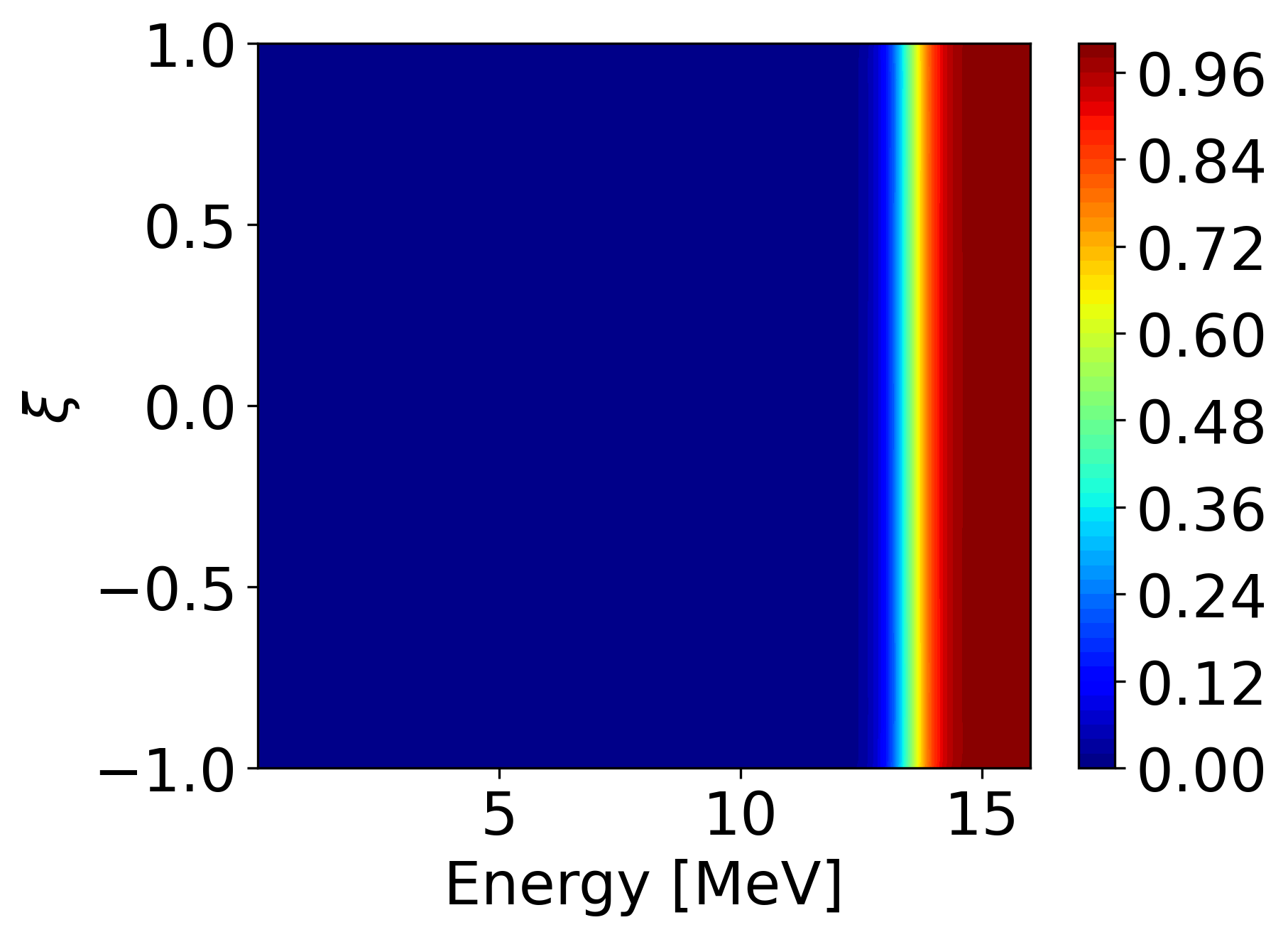}}
    \subfigure[]{\includegraphics[scale=0.33]{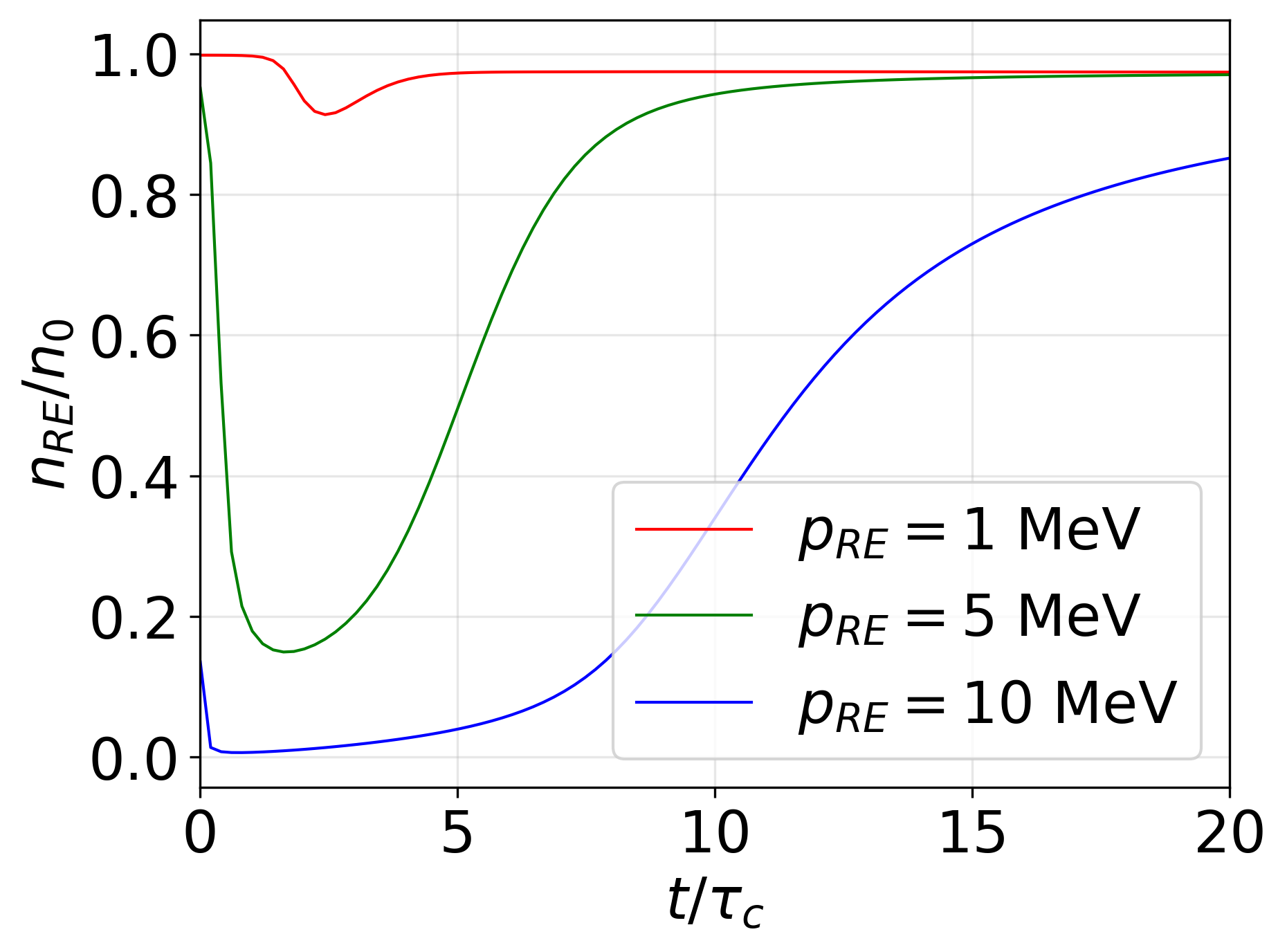}}
    \subfigure[]{\includegraphics[scale=0.3]{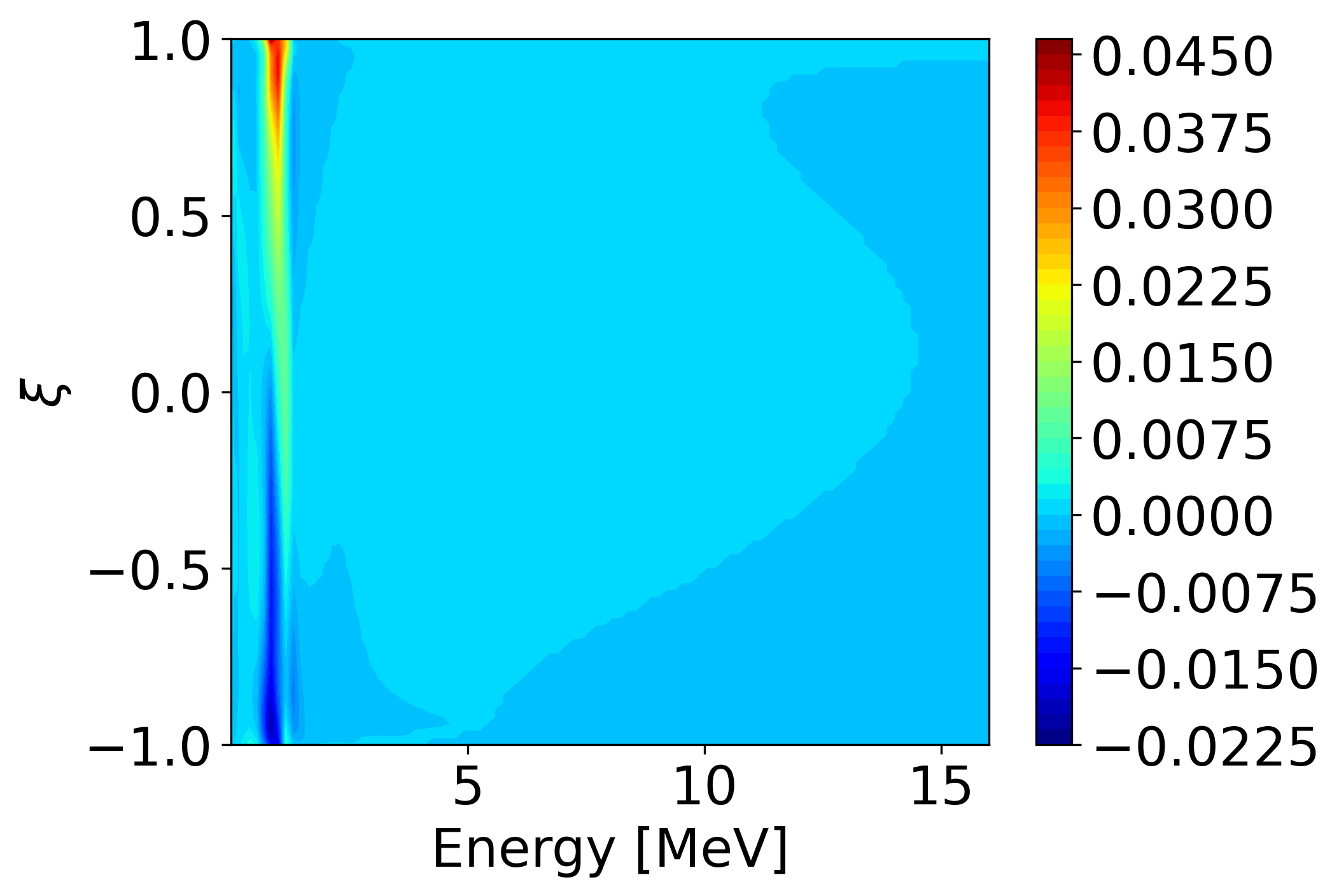}}
    \subfigure[]{\includegraphics[scale=0.3]{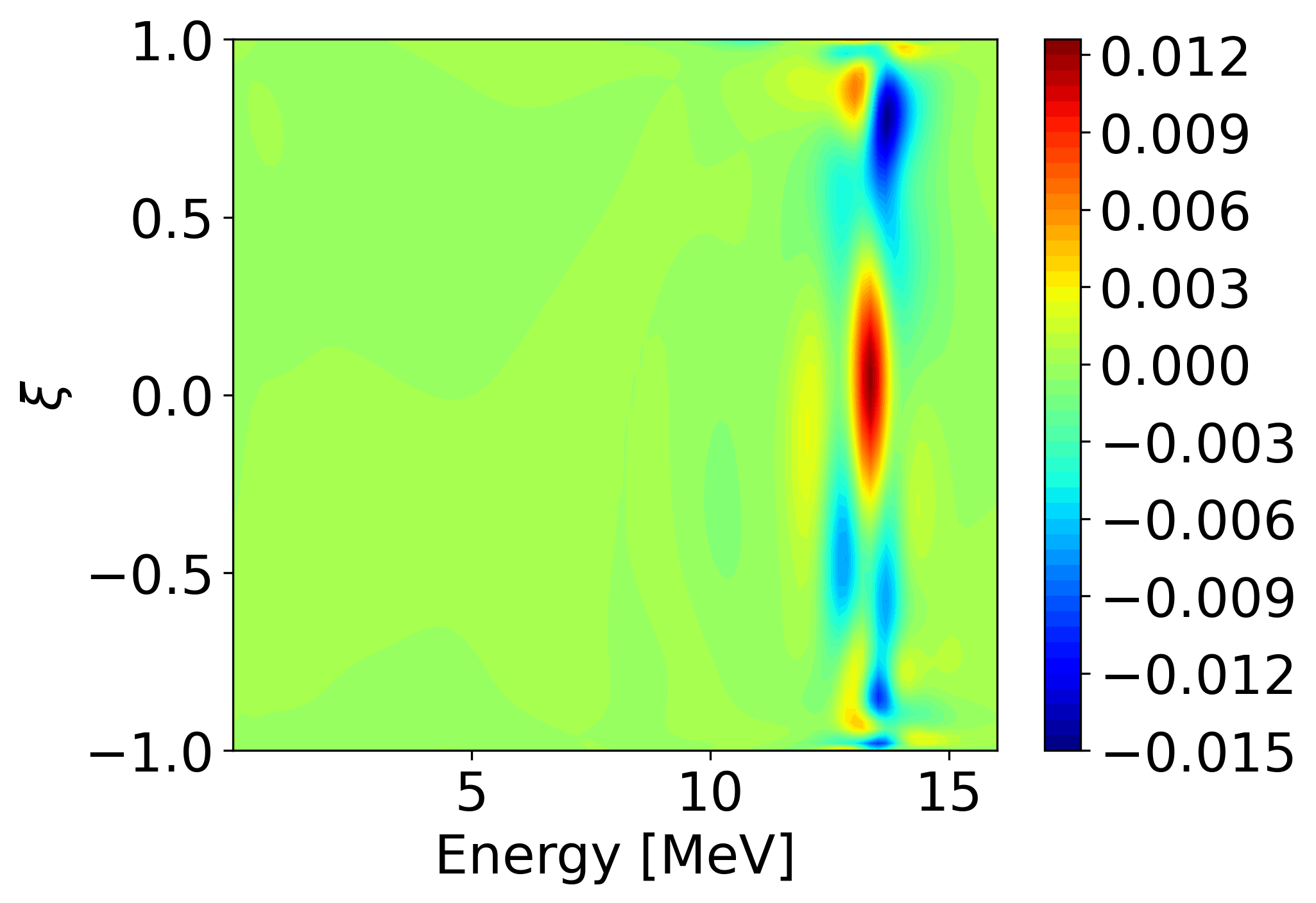}}
    \subfigure[]{\includegraphics[scale=0.35]{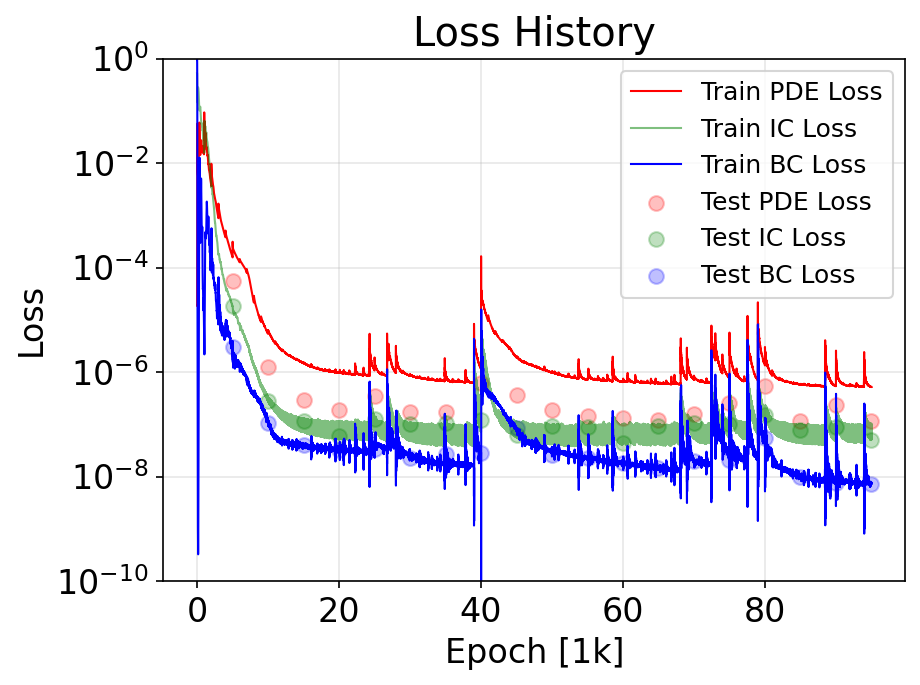}}
    \caption{Panels (a) and (d) show a sharp Heaviside terminal condition at $p_{RE,\min}$ and the residual of the PINN respectively; panels (b) and (e) show a smooth terminal condition at $p_{RE,\max}$ and the residual of the PINN respectively; panel (c) calculates the time dependent RE density computed from Eq.~(\ref{eq:TAS5}) with an isotropic initial condition and an energy distribution discussed in \ref{sec:JONTA} plotted with select variations of $p_{RE}$; panel (f) shows the loss history.}
    \label{energy_distribution_figs}
\end{figure}

This section will utilize the framework described in Sec. \ref{sec:adjointFP} and the algorithm in Sec. \ref{sec:PINNs}
to describe the energy distribution of a RE population under various plasma parameters and initial conditions.
Runaways at low energy typically have sharper features in the energy distribution characterized by the dominant collisional drag, which scales with $1/p^2$,
whereas runaways at high energy have smoother features due to a combination of dynamics in which 
runaways accelerate and become more aligned with the magnetic axis due to electric field pinching, followed by a pitch angle scattering event which places runaways in a decelerating region ($U_p<0$), followed by a rapid deceleration due to synchrotron radiation until
the electric field pinching places runaways in an accelerating region ($U_p>0$) and the cycle continues \cite{decker_hirvijoki_embreus_peysson_stahl_pusztai_fulop_2016, guo_mcdevitt_tang_2017}. 
We choose to capture the varying sharpness in energy by parameterizing a Heaviside transition,
\begin{equation}
    P(\tau=0)=\frac{1}{2}\left[1-\tanh{\left(N(p_{RE})\frac{p_{RE}-p}{p_{\max}-p_{\min}}\right)}\right],
\end{equation} 
where $p_{RE}$ indicates the energy above which an electron is treated as a runaway.
At the lowest $p_{RE}$ we take $N(p_{RE_{\min}})=N_0=100$ and at the highest $p_{RE}$, $N(p_{RE_{\max}})=N_1=30$
and create a continuous formula for the sharpness,
\begin{equation}
    \bar{p}_{RE}= \frac{p_{RE}-p_{RE_{\min}}}{p_{RE_{\max}}-p_{RE_{\min}}} \indent N (p_{RE_{\max}}) = \frac{N_0N_1}{N_1+\left(N_0-N_1\right)\bar{p}_{RE}^2}
\end{equation}
where $\bar{p}_{RE}$ is normalized to the inputs of the PINN where $p_{RE_{\min}}\approx 2.78$ or $1$ MeV and $p_{RE_{\max}}=0.85 p_{\max} \approx 27.6$  or $13.5$ MeV.
An output transform is implemented to enforce that the adjoint solution is bounded between zero and one,
\begin{equation}
    P = \tanh\left[\left(\frac{p-p_{\min}}{p_{\max}-p_{\min}}\right)P_{NN}^2 \right],
\end{equation}
where $P_{NN}$ are the outputs of the hidden layers of the neural network.
We also choose to weight the PDE residual to the synchrotron strength in order to balance the enforced $p_{\max}$ boundary condition with the PDE loss,
\begin{equation}
    L_{PDE} = \frac{1}{N_{PDE}}\sum_{i=1}^{N_{PDE}}\left[\frac{p^2}{1+p^2}\mathcal{R}_i\frac{\alpha_{\min}}{\alpha}\right]^2.
\end{equation}
The energy distribution PINN was trained for 100 thousand epochs with adaptive sampling performed every 1000 epochs where the PINN reached a PDE test loss of $1.15 \times 10^{-7}$, 
the terminal condition test loss reached $5.08 \times 10^{-8}$ and the boundary condition test loss reached $7.28 \times 10^{-9}$.
We note that the spikes in the loss history in Fig \ref{energy_distribution_figs} (f) are due to the adaptive 
sampling routine in which the highest residual is located below threshold regimes near the high energy boundary condition. 


\begin{figure}
    \subfigure[]{\includegraphics[scale=0.33]{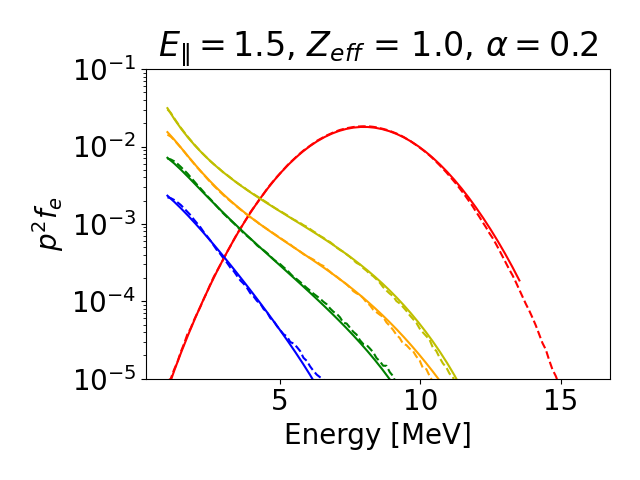}}
    \subfigure[]{\includegraphics[scale=0.33]{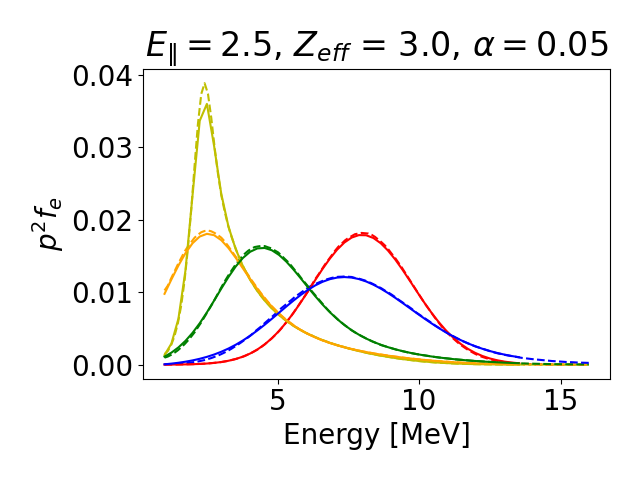}}
    \subfigure[]{\includegraphics[scale=0.33]{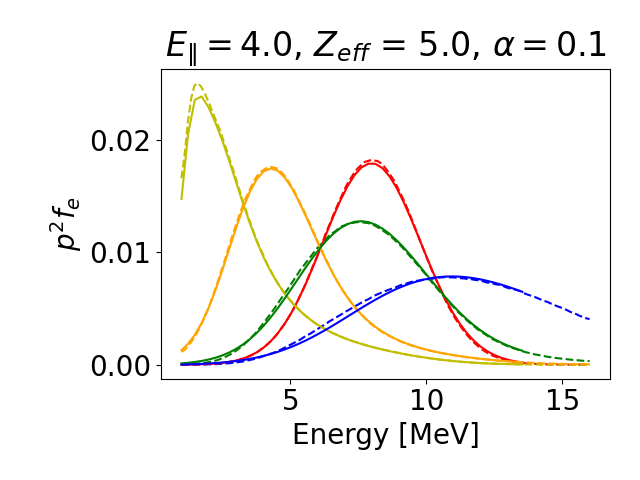}}
    \caption{Panels (a)--(c): Runaway energy distributions for a gaussian and isotropic initialization discussed in \ref{sec:JONTA} across several chosen combinations of plasma parameters, adjoint-PINN [solid] with Monte Carlo [dashed]. Distributions plotted at select times $t=(0,2,5,10,20) \tau_c$ and colors (red, yellow, orange, green, blue) respectively. Panel (a) shows an under threshold case where runaways decelerate; panel (b) shows moderate parameters where runaways saturate; panel (c) shows runaways accelerating beyond the domain.}
    \label{energy_distribution_results}
\end{figure}

The results in Fig. \ref{energy_distribution_figs}(c) demonstrate that a single PINN model, with $p_{RE}$ as an input, can capture the runaway density 
above $p_{RE}$ described in Eq. (\ref{eq:TAS5}) whereby 
the results in Fig. \ref{energy_distribution_results} demonstrate that this model can likewise capture the energy distribution via Eq. (\ref{Adjoint energy distribution}).
This PINN model is capable of capturing orders of magnitude variations in the energy distribution as well as the acceleration and deceleration of runaways through the boundaries.
Panel (a) shows a combination of plasma parameters under the threshold electric field ($E_{\Vert,\mathrm{thresh}}\approx 1.66$) in which runaways 
undergo a constant decay to the low energy bulk distribution.
Panel (b) shows a combination of plasma parameters slightly above threshold ($E_{\Vert,\mathrm{thresh}}\approx 1.61$) in which isotropic runaways decelerate to 
low energy, with some loss of electrons, and reaccelerates until the energy distribution saturates and becomes steady state.
Panel (c) shows a combination of plasma parameters far above threshold ($E_{\Vert, \mathrm{thresh}}\approx 2.11$) in which runaways partially escape the domain. 

\FloatBarrier
\section{Discussion and Conclusion}
\label{sec:discussion_conclusion}
This work describes a novel neural network framework for runaway electron kinetic surrogates 
where we present a technique for predicting any fluid moment and the energy distribution of runaways.
Through the utilization of physics-informed neural networks (PINNs), this framework is capable of scanning the parametric dependence of a wide range of plasma parameters.
While computationally intensive to train, this powerful framework enables nearly instantaneous prediction of runaway moments and the energy distribution for arbitrary initial conditions, as well as orders of magnitude larger time steps than with traditional methods.



Fluid moments are traditionally found through the fully kinetic distribution which is computationally intensive for multi-physics coupling or via reduced-order fluid approximations which reduce fidelity. 
We present an adjoint framework for accurately capturing fluid moments, with kinetic fidelity, for regimes in which runaways are well within the energy domain, however, for regimes in which runaways escape the domain, the adjoint method can subtly diverge. 
For example, the current moment is accurately captured due to the velocity weighting of the moments which negligibly changes at relativistic energies.
The energy moment, however, can diverge well above threshold such that runaways which escape through the high energy boundary are assumed to be frozen at the boundary which is untrue.
We mitigate this description of the energy moment via a higher fidelity description of runaway energies through efficient computation of the runaway energy distribution.

The energy distribution is critical for diagnosing and mitigating tokamak disruptions and 
addresses the discussed energy moment by capturing the proportion of runaways which exit the domain.
This is accomplished via careful treatment of the adjoint problem by defining runaway electrons to be within a narrow energy bin described in Sec. \ref{sec:energy_distribution1} whereby this energy bin is parameterized as an input to the PINN.
We show that the adjoint-PINN framework can accurately predict orders of magnitude variation of the energy distribution as well as scan a wide range of plasma parameters to capture the acceleration and deceleration beyond the domain.

Future work includes utilizing this framework towards predicting higher-order moments such as the pressure and heat flux moment as well as extending the energy domain, through proper normalizations, to predict runaway dynamics at even higher relativistic energies.
Further, we intend to extend this framework to a broader range of plasma parameters, as well as include more accurate collisional frequencies including partial screening \cite{hesslow_thesis,hesslow_embreus_stahl_dubois_papp_newton_fulop_2017,hesslow_embreus_hoppe_dubois_papp_rahm_fulop_2018,arnaud2025runaway}.

\begin{acknowledgements}

This work was supported by the Department of Energy, Office of Fusion Energy Sciences at the University of Florida under awards DE-SC0024649 and DE-SC0024634. The authors acknowledge the University of Florida Research Computing for providing computational resources that have contributed to the research results reported in this publication.

\end{acknowledgements}

\bibliographystyle{apsrev}
\bibliography{energy_reconstruction_manuscript2.bib}

\end{document}